\documentstyle[aps,pre,twocolumn,graphicx,psfig]{revtex}

\begin{document}
\draft

\author{K. Dawson$^1$, G. Foffi$^1$, M. Fuchs$^2$, W. G\"otze$^2$, F. Sciortino$^3$, M. Sperl$^2$, P. Tartaglia$^3$, Th. Voigtmann$^2$, E. Zaccarelli$^1$}
\address{$^1$Irish Centre for Colloid Science and Biomaterials, University College Dublin, Belfield, Dublin 4, Ireland\\
$^2$Physik-Department, Technische Universit{\"a}t M{\"u}nchen, D-85747 Garching, Germany\\
$^3$Dipartimento di Fisica and Istituto Nazionale per la Fisica della Materia, Universit\`{a} di Roma La Sapienza, P.le A. Moro 2, I-00185 Roma, Italy}

\title{Higher order glass-transition singularities 
in colloidal systems with attractive interactions.}

%\date{received 20 June 2000, accepted 24 August 2000}
\date{Phys. Rev. E, in print }

\maketitle

\begin{abstract}
The transition from a liquid to a glass in colloidal suspensions
of particles interacting through a hard core plus an attractive
square-well potential is studied within the mode-coupling-theory framework.
When the width of the attractive potential is much shorter than the hard-core
diameter, a reentrant behavior of the liquid-glass line, and a glass-glass-transition
line are found in the temperature-density plane of the model. For small well-width values,
the glass-glass-transition line terminates in a third order bifurcation point, i.e.\ in a
$A_3$ (cusp) singularity. On increasing the square-well width, the glass-glass
line disappears, giving rise to a fourth order $A_4$ (swallow-tail) singularity at a
critical well width. Close to the $A_3$ and $A_4$ singularities the decay of the
density correlators shows stretching of huge dynamical windows, in
particular logarithmic time dependence.
\end{abstract}


\section{Introduction}
\label{sec:intro}

Colloidal suspensions have been studied extensively because of 
their practical importance and because of their relevance in biophysics. These
systems are also of great theoretical interest since they are models for 
conventional matter. They can be prepared for a large span of densities so that 
the states can be gases, gels, liquids, crystalline solids, or 
glasses. Light scattering can be used to measure the static
structure factor and various correlation functions. The dynamics
can be explored over a wide range of length scales and over huge dynamical 
windows \cite{Russel89,Pusey91}. Fascinating with colloidal 
systems is that the interaction can be tuned to some extent by varying the
coating of the particles and the composition of the solvent 
\cite{Russel89,Pusey91,Poon98}. It is possible to realize the hard-sphere
system (HSS), the basic model underlying all theories of simple 
liquids \cite{Hansen86,Balucani94}. One can also prepare systems where
the hard core is complemented by an attractive shell.
This allows to study the interplay of repulsion and attraction.
As a contribution to such studies, a theory for the glass formation
resulting from a strong short-range attraction among densely packed
hard-sphere colloidal particles shall be presented in this paper.

In hard-sphere colloidal dispersions,
the liquid-glass transition has been studied by van~Megen and Pusey \cite{Megen91}.
They measured correlation functions $\phi_q(t)$ for density fluctuations
of a representative set of wave numbers $q$ over about four decades in time $t$.
It was found that these correlations decay to zero as expected for a liquid only
for packing fractions $\varphi$ below a critical value $\varphi_c$. At $\varphi_c$,
the long-time limit of the correlators, $f_q=\phi_q(t\rightarrow\infty)$, changes
discontinuously to a certain value $f_q^c>0$, increasing further with
packing fraction. $f_q$ is the Debye-Waller factor of the amorphous solid, i.e.\ of
the glass, and generalizes the order parameter
introduced by Edwards and Anderson in the theory of spin glasses \cite{EA75}.
The evolution of the glassy dynamics for the HSS
was studied comprehensively by van~Megen and coworkers
\cite{Megen91b,Megen93,Megen93b,Megen94,Megen94b,Megen98}. The data suggest,
that it is the well known cage effect \cite{Balucani94} which causes the glassy
dynamics and the arrest of density fluctuations at $\varphi_c$.

The cage effect is the essential physical concept underlying the mode-coupling
theory (MCT) for the evolution of glassy dynamics in simple systems 
\cite{Bengtzelius84,Leutheusser84}. This theory allows the calculation of 
$\phi_q(t)$ and thus $f_q$ from the equilibrium structure factor $S_q$.
As a function of control parameters like $\varphi$, singularities of $f_q$,
called glass-transition singularities, may occur. The simplest type, called a fold
bifurcation, describes a liquid-glass transition at
$\varphi=\varphi_c$. It implies a subtle dynamical scenario,
giving rise to universal features of glassy dynamics, which have been identified
in a leading-order-asymptotic expansion of the MCT equations.
A review of the basic results is given in Ref.~\cite{Goetze92}.
In Refs.~\cite{Megen91b,Megen93,Megen93b,Megen94,Megen94b}, detailed 
quantitative comparisons of the data for hard-sphere colloids with the MCT
predictions are presented. It is shown that the theory accounts
for the experimental facts within a 15\%-accuracy level. An illuminating 
summary of these studies is given in Ref.~\cite{Megen95}. Results for
the shear modulus have also been interpreted with the universal MCT formulas \cite{Mason95}.
The evolution of glassy dynamics for $\varphi$ increasing towards $\varphi_c$
was also studied for polymer micronetwork colloids 
\cite{Bartsch92,Bartsch93,Bartsch95b,Bartsch97}.
Here, the interparticle interaction is not known. But the authors
demonstrated, that a consistent fit of their data with the universal 
MCT formulas was possible.
Preliminary studies of the glassy dynamics of charge-stabilized colloids 
indicate, that these data can also be explained within MCT \cite{Beck99}.
The reported findings shall be taken as a justification to base the theory
in this paper on the MCT for simple systems.

Our studies deal with the square-well system (SWS), characterized
by a hard-core repulsion for interparticle distances $r<d$, and by a constant
attraction potential within the shell $d<r<d+\Delta$. The theory focuses on the
high-density regime, say $\varphi>0.4$, so that the cage effect is essential
for the dynamics. The relative attraction-shell width
$\delta=\Delta/d$ is assumed to be small, say $\delta<0.15$. The main
outcome of our theory is the prediction of a higher-order glass-transition
singularity at a critical packing fraction $\varphi^{*}$ somewhat above the 
critical point $\varphi_c$ of the hard-sphere system and a critical width
$\delta^{*}$ of about $0.04$. This singularity organizes a subtle phase diagram
and opens up various possibilities for glassy relaxation. The new results reflect the 
interplay of two mechanisms for particle localization, i.e. for the arrest of
density fluctuations. It can either be dominated by repulsion of the 
particle by its cage-forming neighbors, or by the formation
of bonds to the boundaries of the cage. Preliminary calculations 
\cite{Fabbian99,Bergenholtz99} based upon Baxter's adhesive-hard-sphere model
\cite{baxter2} hinted at some findings to be derived in this paper. Baxter's
model treats the limit $\delta\rightarrow 0$, so it cannot deal with
the indicated singularity at $\delta^{*}$. Moreover, taken literally, the 
Baxter model cannot be used as a basis of MCT applications, since there appears 
a divergency due to excitations with large wave vectors. The results for this
model in Refs.~\cite{Fabbian99,Bergenholtz99,Foffi00} depend in an ill-defined manner on
the large-$q$ cutoff used there, a problem which is avoided with the SWS.

Dense systems of colloidal particles characterized by a hard core and
strong attractions of a range smaller than the core diameter by a factor of at
least 10 were realized experimentally, when adding nonadsorbing polymers to
either a suspension of colloidal hard spheres \cite{Poon93} or to emulsions
\cite{Meller99}, in solutions of sterically stabilized particles when
decreasing the solvent quality \cite{Grant93,Verduin95,Rueb97,Rueb98}, 
and in copolymer micellar systems when changing the temperature \cite{Lobry99}.
Such systems were also studied in Monte Carlo simulations 
\cite{Kranendonk88,Lang99}. Non-equilibrium phenomena characterized by a 
number of aspects were found which cannot be understood from the glassy states 
formed in hard-sphere solutions. 
First, amorphous solids could be formed by increasing the attraction strength
even though the packing fraction was kept fixed well below the value of the
hard sphere glass transition \cite{Meller99,Grant93,Verduin95,Rueb97,Rueb98}.
Second, increasing the strength of a short-ranged attraction by  adding small
polymers, melting of the glass states was reported for the colloid-polymer 
mixtures \cite{Lekkerkerker92,Poon93}. 
Third, the non-decaying frozen structures seen when immersing polymer coated 
colloidal particles into  solvents of decreasing quality \cite{Verduin95} 
exhibited a  much larger Debye-Waller factor at small wave vectors than  hard-sphere 
systems. This indicates a much higher rigidity of the solid states on 
intermediate length scales. In support of this observation, viscoelastic 
measurements for intermediate frequencies found strongly concentration 
dependent elastic moduli \cite{Meller99,Grant93,Rueb97,Rueb98}. 
It will be shown that our results provide a qualitative explanation of the 
reported findings.

The paper is organized as follows. In Sec.~\ref{sec:sofq}, we report our results
for the structure factor of the SWS and discuss those features which cause 
various qualitative results of the MCT solutions. Section~\ref{sec:phase}
presents the main result of this paper, showing the phase diagram, and discussing
the properties of the glass states resulting from the interplay between attraction
and repulsion. In Sec.~\ref{sec:dynamics}, we present some results for the dynamics
which illustrate that the higher order glass-transition singularities cause relaxation
stretching which is much more pronounced than is known for the HSS. The last
Sec.\ref{sec:conc} presents some concluding remarks.

\section{Structure Factor Calculations}
\label{sec:sofq}

\subsection{The Model}
\label{sec:sofq:model}

The structure factor $S_q$ is the essential input information needed
to formulate the MCT equations. In this section, $S_q$ shall be 
discussed for the square-well system (SWS). Only such states shall be considered
for which $S_q$ depends smoothly on the particle density $\rho$, on the
temperature $T$, and on the wave number $q$. The interaction potential $V(r)$
for particles with separation distance $r$ consists of a hard-core repulsion 
for $r<d$, and it has the negative value $-u_0$ within the attraction shell
$d<r<d+\Delta$. The structure can be specified by three control parameters: 
the packing fraction $\varphi$ of the hard cores, the ratio $\theta$ of 
thermal and attractive energy, and the relative width $\delta$ of the 
attraction shell:
\begin{equation} 
\label{eq:param_ext}
\varphi = \pi \rho d^3/6,\quad
\theta  = k_B T/u_0,\quad
\delta	= \Delta/d\, .
\end{equation}

Let us note the standard concepts needed for a discussion of $S_q$ 
\cite{Hansen86}. $g(r)$ and $h(r)=g(r)-1$ abbreviate the pair distribution and 
the total correlation function, respectively. The Fourier transform $h_q$ of
the latter determines the structure factor $S_q=1+\rho h_q$.
The Ornstein-Zernike equation formulates an integral equation for $h(r)$,
where the kernel is the direct correlation function $c(r)$. In the
wave vector domain, it reads $S_q=1/[1-\rho c_q]$, where

\begin{equation} 
\label{eq:cq}
c_q = \frac{4\pi}{q} \int_0^\infty {d}r \sin(q r) [r c(r)]\,.
\end{equation}
Baxter's method of the Wiener-Hopf factorization \cite{Hansen86,Baxter68} shall be
used to reformulate the Ornstein-Zernike equation. The basic concept of this theory
is the factor function $Q(r)$. It is defined as a continuous real function for $r\ge0$,
determining $S_q$ via its Fourier transform:

\begin{mathletters}\label{eq:Sq}
\begin{eqnarray}\label{eq:Sq-a}
S_q^{-1}   &=& \hat{Q}(q)\hat{Q}(q)^*\,,\\\label{eq:Sq-b}
\hat{Q}(q) &=& 1 - 2 \pi \rho \int_0^\infty {d}r \exp(i q r) Q(r)\,.
\end{eqnarray}\end{mathletters}
It is anticipated that $Q(r)$ as well as $c(r)$ vanishes beyond a certain 
distance $R$. For $0\le r\le R$, there holds
\begin{equation}\label{eq:WH1-b}
r c(r) = - Q'(r) + 2 \pi \rho \int_r^R {d}s\, Q'(s)Q(s-r)\,.
\end{equation}
Furthermore, one finds for $r>0$
\begin{equation}\label{eq:WH2-a}
r h(r) = - Q'(r) + 2 \pi \rho \int_0^R {d}s (r-s) h(|r-s|) Q(s)\,.
\end{equation}

For the SWS, $g(r)=0$ is fulfilled for $0<r<d$, and therefore, using $h(r)=g(r)-1$,
Eq.~(\ref{eq:WH2-a})
splits into three subequations. Most simple is the result for the middle part,
$\Delta\le r \le d$, where the formula known from the theory for the 
hard-sphere system (HSS) is reproduced:
\begin{mathletters}\label{eq:WHQ}
\begin{equation}\label{eq:WHQ-a}
Q'(r) = a r + b\,.
\end{equation}
Here, coefficients $a$ and $b$ are introduced by
\begin{equation}\label{eq:WH:ab}
a = 1 - 2 \pi \rho \int_0^{d+\Delta} {d}s \,Q(s)\,,
\quad
b=2 \pi \rho \int_0^{d+\Delta} {d}s \,s\, Q(s)\,.
\end{equation}
Writing $G(r)=rg(r)$, one finds for small distances, $0\le r\le \Delta$,
\begin{equation}\label{eq:WHQ-b}
Q'(r) = a r + b - 2\pi\rho \int_{r+d}^{d+\Delta}{d}s\,G(s-r)Q(s)\,,
\end{equation}
and for the attraction shell, $d\le r\le d+\Delta$, one obtains
\begin{equation}\label{eq:WHQ-c}
Q'(r) = a r + b -G(r) + 2\pi\rho \int_{0}^{r-d}{d}s\,G(r-s)Q(s)\,.
\end{equation}
\end{mathletters}

Some approximation for $c(r)$ has to be introduced into Eq.~(\ref{eq:WH1-b})
in order to close the system of Eqs.~(\ref{eq:WH1-b}) and (\ref{eq:WHQ}).
In this paper, the Percus-Yevick approximation (PYA) and the mean-spherical
approximation (MSA) shall be applied \cite{Hansen86}.
Nezbeda already studied the structure factor for the SWS using the PYA for small
well widths \cite{Nezbeda74,Nezbeda77}. His equations could be solved
only in a restricted region of parameters. Since the boundary of this region of
applicability is close to the parameter region $\varphi \approx 0.5,\,
\theta\approx 1$ of interest in this paper, it does not seem appropriate
to base the following calculations on these results.

\subsection{Approximations}
\label{subsec:sofq:approx}

Within the PYA, one writes $c(r) = g(r) [1-\exp(V(r)/k_B T)]$
outside the hard core.
Substitution of this ansatz into Eq.~(\ref{eq:WH1-b}) and using 
Eq.~(\ref{eq:WHQ-c}) leads to the approximation for 
$d\le r \le d+\Delta$
\begin{eqnarray}\label{eq:PYA}
e^{-u_0/k_B T} G(r) =&& a r + b - 2\pi\rho\int_r^{d+\Delta}{d}s\,
			  Q'(s) Q(s-r)\nonumber\\
		&&+ 2\pi\rho\int_0^{r-d}{d}s\,G(r-s) Q(s)\,.
\end{eqnarray}
Equations (\ref{eq:WHQ}) and (\ref{eq:PYA})
for $Q(r)$ and $G(r)$ are solved numerically. To proceed, the equations 
are discretized straightforwardly. On each of the three $r$-intervals,
a grid of equally spaced points $r_n$ is chosen, where $n=1,2,\dots,1000$.
The functions $Q'(r)$ and $G(r)$ are calculated iterating Eqs.~(\ref{eq:WHQ})
and (\ref{eq:PYA}). At each step, the function $Q(r)$ is evaluated from $Q'(r)$
using a 5-point numerical integration. The procedure is carried out until the
difference between two successive iterations summed over all points of the
$r$-grid becomes less than $10^{-12}$.
The integral in Eq.~(\ref{eq:Sq-b}) is determined by a simplified Filon 
procedure to obtain $\hat{Q}(q)$ and hence $S_q$.

The MSA uses $c(r) = -V(r)/k_B T$ outside the hard core.
Substituting this ansatz into Eq.~(\ref{eq:WH1-b}), after integration one obtains
for $d \le r \le d + \Delta$
\begin{eqnarray}\label{eq:MSA}
Q(r)&=&2\pi\rho\int_{r}^{d+\Delta}{d}s\, Q(s) Q(s-r)\nonumber\\
  &&+ \left[(d+\Delta)^2-r^2\right]/(2\theta)\;.
\end{eqnarray}
Equations (\ref{eq:WHQ}) and (\ref{eq:MSA})
are solved analytically in a leading and next-to-leading order expansion,
using the well width $\delta$ as the small parameter. For the organization of
the expansion, the quantity $K = \delta/\theta$ is considered fixed.
This procedure is motivated by Baxter's theory of sticky hard spheres 
\cite{baxter2}. He evaluated $S_q$ in the limit $\delta\rightarrow 0,\,
u_0\rightarrow\infty$, keeping a parameter equivalent to $K$ fixed.
Details of the calculation can be found in the appendix. The hard core 
diameter $d$ shall be used as the unit of length. For
$\delta\le r\le1$, the factor function is the parabola known from 
the theory of the HSS:
\begin{mathletters}\label{eq:SWF:Qr}
\begin{equation}\label{eq:SWF:Qr-a}
Q(r) = a r^2/2 + b r + c\,,
\end{equation}
with $a$, $b$, and $c$ now being smooth functions of the SWS control parameters.
For $0\le r\le\delta$, there is an enhancement above this parabola:
\begin{equation}\label{eq:SWF:Qr-b}
Q(r) = a r^2/2 + b r + c + 2 \varphi K^2\cdot 
\delta \left[1-(r/\delta)\right]^3\,.
\end{equation}
Within the attraction shell, the leading order result describes a linear 
decrease of $Q(r)$ from $K$ to zero. The leading correction adds 
a quadratic modification. One finds for $1\le r\le 1+\delta$
\begin{eqnarray}\label{eq:SWF:Qr-c}
Q(r) &= & K \left[1-\frac{r-1}{\delta}\right] 
\nonumber\\
&+&K\delta\cdot\left\{
\frac{1}{2}\left[1-\left(\frac{r-1}{\delta}\right)^2\right]
+ 6 \varphi c_0 \left[1-\frac{r-1}{\delta}\right]^2
\right\}\,.
\end{eqnarray}
\end{mathletters}
Here and in the following we denote the constants from Eq.~(\ref{eq:SWF:Qr}) as
$a = a_0 + K\delta\cdot a_1,\,b = b_0 + K\delta\cdot b_1,$ and
$c = c_0 + K\delta\cdot c_1$. 
The leading
order contributions are the result of the Baxter limit $\delta\rightarrow 0$:
\begin{mathletters}\label{eq:SWF:abc0}
\begin{eqnarray}\label{eq:SWF:abc0-a}
a_0 &=& \left[\frac{1+2\varphi}{(1-\varphi)^2}\right]
- \frac{12 K \varphi}{(1-\varphi)}\,,
\\\label{eq:SWF:abc0-b}
b_0 &=& \left[\frac{-3 \varphi}{2(1-\varphi)^2}\right] 
+ \frac{6 K \varphi}{(1-\varphi)}\,,
\\\label{eq:SWF:abc0-c}
c_0 &=& \left[\frac{-1}{2(1-\varphi)}\right] 
+ K\,.
\end{eqnarray}
\end{mathletters}
The terms in brackets exhibit the results for the HSS \cite{Hansen86}.
The coefficients of the next-to-leading-order contributions are
\begin{mathletters}\label{eq:SWF:abc1}
\begin{eqnarray}\label{eq:SWF:abc1-a}
a_1 &=& \left[6\varphi(5\varphi-2)-72 c_0 \varphi^2(1-\varphi)\right]/(1-\varphi)^2
\,,
\\\label{eq:SWF:abc1-b}
b_1 &=& \left[9\varphi(1-2\varphi)+36 c_0 \varphi^2(1-\varphi)\right]/(1-\varphi)^2
\,,
\\\label{eq:SWF:abc1-c}
c_1 &=& \left[1-7\varphi+12 c_0 \varphi(1-\varphi)\right]/(2(1-\varphi))\,.
\end{eqnarray}
\end{mathletters}

Substitution of Eq.~(\ref{eq:SWF:Qr}) into Eq.~(\ref{eq:Sq-b}) yields
$\hat{Q}(q)$ as a combination of trigonometric functions. It is elementary
to work out the somewhat lengthy expression
and thus via  Eq.~(\ref{eq:Sq-a}) the desired result for $S_q$.

The large-$q$ asymptote of the direct correlation function $c_q$ shall be obtained
from Eq.~(\ref{eq:cq}) by evaluating the asymptote of the Fourier-sine transform of
the function $f(r)=r c(r)$.
From Eq.~(\ref{eq:WH1-b}) one concludes, that $f(r)$ is smooth except for
at most three points $r^{(1)}=\delta,\,r^{(2)}=1,$ and $r^{(3)}=1+\delta$.
At these points there can be a discontinuity, given by that of the derivative
of the factor function: $f^{(i)} = Q'(r^{(i)}-0)-Q'(r^{(i)}+0)$.
Let us also note from Eq.~(\ref{eq:WH1-b}) the initial value
$f(r=0)=A=-Q'(0)-6\varphi Q(0)^2$. For the exact solution of the problem, $A=0$
must hold, but due to the approximation scheme used here, a finite value of
${\mathcal O}(\delta)$ remains: $A=-K\delta\left[12\varphi c_0(c_1+2K\varphi)+b_1\right]
+{\mathcal O}(\delta^2)$. Thereby, the Baxter result \cite{baxter2},
$A={\mathcal O}(\delta^0)$, is improved. The $f^{(i)}$ can be
determined easily from Eq.~(\ref{eq:SWF:Qr}), in particular $f^{(1)}=0$.
Integrating by parts, the integral in  Eq.~(\ref{eq:cq}) becomes:
$\left[f^{(0)}+\sum_i f^{(i)} \cos(q r^{(i)})\right]/q + {\mathcal O}(1/q^2)$.
Hence one arrives at $c_q = c_q^{\text{as}} + {\mathcal O}(1/q^3)$,
where the asymptotic tail reads
\begin{eqnarray}\label{eq:cq_as}
c_q^{\text{as}} = (4\pi/q^2)&\cdot&\left[A + B\cos(q) \right.
\nonumber\\
 &&+\left.
  2 C \sin\left(q (1+\delta/2)\right)\sin\left(q \delta/2\right)
\right]\,.
\end{eqnarray}
The second term in the bracket has a form familiar from the PYA result
for the HSS. But the coefficient $B$ is a smooth function of $K$ and
$\delta$ which reduces to the HSS value for $K=\delta=0$:
\begin{mathletters}\label{eq:SWF:ABC}
\begin{equation}\label{eq:SWF:ABC-a}
B = a + b + K(12\varphi c_0 - 1)\,.
\end{equation}
The third term in the bracket is due to the existence of the attraction well.
Its prefactor reads
\begin{equation}\label{eq:SWF:ABC-c}
C = (1+\delta)/\theta\,.
\end{equation}
\end{mathletters}

\subsection{Results}
\label{subsec:sofq:results}
\begin{figure}
\centerline{\scalebox{0.75}{\includegraphics{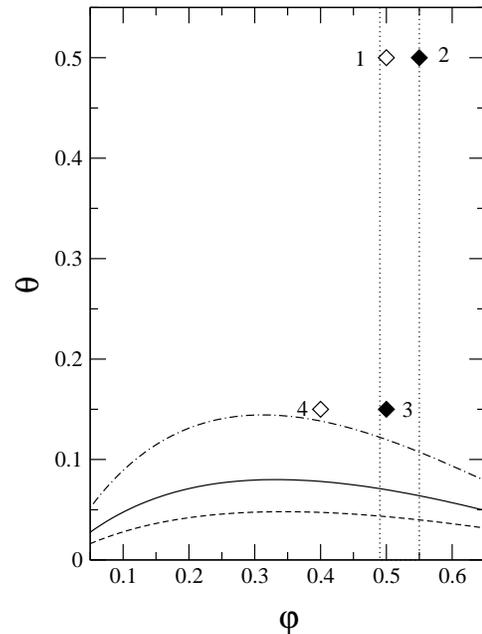}}}
\caption{Control-parameter plane for the square-well system (SWS), plotted as
dimensionless temperature $\theta = k_B T / u_0$ versus packing fraction
$\varphi$. The full line shows the spinodal calculated
within the MSA for the relative attraction-well width $\delta=0.05$.
Dashed (dash-dotted) curves show the corresponding spinodals for
$\delta=0.03$ ($\delta=0.09$).
Vertical dotted lines mark the region for which the phase diagram is 
discussed below in Fig.~\ref{fig:fig5}. Diamonds mark the state parameters
for which the structure factor is shown in  Fig.~\ref{fig:fig2}.
\label{fig:fig1}}
\end{figure}

The spinodal lines of the SWS are shown in Fig.~\ref{fig:fig1}
for three representative values of the well width $\delta$.
They specify the divergence points of the compressibility, i.e. the zeros
of $S_q^{-1}$ for $q=0$. 
The spinodal is the boundary of the regime of absolute instability 
with respect to the liquid-vapor transition. Only states outside this regime
can be considered in the following.
Substitution of Eq.~(\ref{eq:SWF:Qr}) into 
Eq.~(\ref{eq:Sq-b}) yields elementary expressions for $\hat{Q}(0)$ within 
MSA. We have not been able to determine the spinodals within 
the PYA, due to numerical instability of the algorithm.
The high-density regime investigated in the following applications of the
MCT is indicated as the strip between the two dotted vertical lines.

\begin{figure}
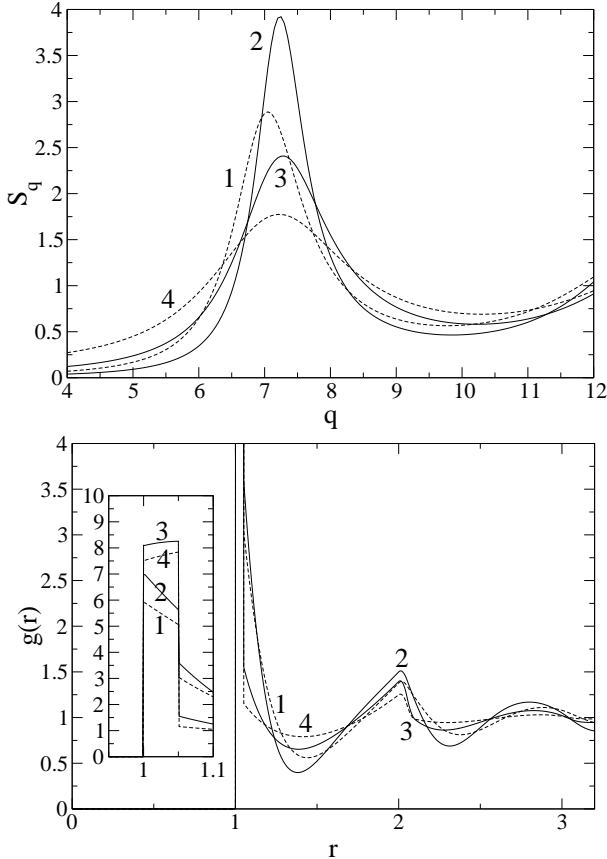

\centerline{\scalebox{0.5}{\includegraphics[angle=-90]{fig2a.eps}}}
\nobreak
\centerline{{\includegraphics[width=.43\textwidth]{fig2b.eps}}}
\caption{Structure factor $S_q$ and pair-correlation function $g(r)$ of the SWS
calculated within the MSA for relative well width $\delta=0.05$. The labels 1 to 4
correspond to the states indicated by the diamonds in Fig.~\ref{fig:fig1}. They are
given by the pairs $(\varphi,\theta)$ of packing fraction and reduced temperature
(0.50, 0.50), (0.55, 0.50), (0.50, 0.15), and (0.40, 0.15), respectively. Here and in
the following figures, the hard core diameter is chosen as the unit of length, $d=1$,
and $q$ is given here and in the following in units of $d^{-1}$.
\label{fig:fig2}}
\end{figure}

Figure~\ref{fig:fig2} exhibits structure factors $S_q$ calculated within the MSA for 
$\delta=0.05$, and the corresponding pair distributions $g(r)$, calculated numerically
from Eq.~(\ref{eq:WH2-a}), for states marked by diamonds in Fig.~\ref{fig:fig1}.
The $S_q$-versus-$q$ curves exhibit a principal refraction peak as known from other
simple liquids \cite{Hansen86}. It is caused by the hard-core driven excluded volume
phenomenon, $g(r<d)=0$. The high-temperature curves 1 and 2 exhibit peaks, which are
only slightly smaller and somewhat broader than the peaks of a HSS at the same densities.
The attraction modifies the pair correlations and thus the excluded volume effects,
as can be inferred by comparing the curves 1 and 3. Lowering $T$, the short-ranged
attraction causes the particles to move closer, i.e.\ the most probable interparticle spacing
decreases. Therefore, the peak position shifts to higher $q$ upon cooling.
The distribution $g(r)$ develops a more rapidly varying structure at distances
which are multiples of the particle diameter, and this explains the decrease of the
peak height and the increase of the peak wings in $S_q$. A change of the density at low 
temperature modifies the peak in a similar manner as discussed above for large
$T$, cf.\ curves 3 and 4. However, lowering
$\varphi$ drives the system closer to the spinodal, and therefore the exhibited
change of $S_q$ for small $q$ is larger than expected for a HSS.

Results for the pair correlation $g(r)$ obtained by different closures of
Eqs.~(\ref{eq:WH1-b}) and (\ref{eq:WH2-a}) and by other methods have been published
by Lang \emph{et~al.} \cite{Lang99}. For both a small and a large well width
considered there, $\delta=0.03$ and $\delta=0.5$ respectively, we find our results
for $g(r)$ in agreement with the Monte-Carlo simulation results obtained by
Lang \emph{et~al.} Only for the small well width, $g(r)$ is underestimated in
the well regime, $1<r<1+\delta$, by about the same amount as Nezbeda's approximation
\cite{Nezbeda74,Nezbeda77} overestimates $g(r)$. At $r>1+\delta$,
however, our solution appears to be in better agreement with the simulation
results. This behaviour is similar to what holds for the optimized
random-phase approximation \cite{Lang99}.

\begin{figure}
\centerline{\scalebox{0.7}{\includegraphics{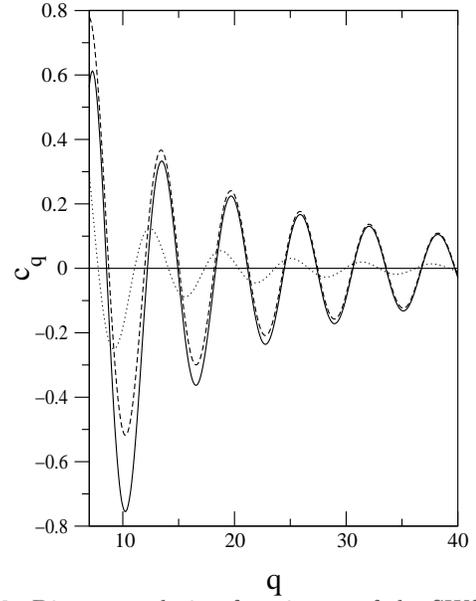}}}
\caption{Direct correlation function $c_q$ of the SWS for relative well width
$\delta = 0.05$, calculated within the MSA (solid line). Density and 
temperature are the ones considered in Figs.~\ref{fig:fig1} and \ref{fig:fig2}
for the label 3. 
The dashed line exhibits the leading asymptote $c^{\text{as}}_q$ according to 
Eq.~(\ref{eq:cq_as}).
The dotted line represents the same result with coefficient $C$ replaced by zero
(see text).
\label{fig:fig3}}
\end{figure}

The large-$q$ tail of $c_q$ will be of importance in the following. In
Fig.~\ref{fig:fig3} it is shown that the
asymptote, Eq.~(\ref{eq:cq_as}), describes $c_q$ very well for $q>20$.
The results have been evaluated for the state discussed in 
Fig.~\ref{fig:fig2} with the label~3, where $(A, B, C) = (-0.092, 1.63, 7)$. 
The tail consists of a
part due to the first two terms in Eq.~(\ref{eq:cq_as}), which differs from
the HSS result merely by modifications of the coefficients $A$ and $B$. This part of the
asymptote is shown in Fig.~\ref{fig:fig3} (dotted line) in order to emphasize
that the last contribution in Eq.~(\ref{eq:cq_as}) can be dominant.
The next-to-leading order contributions to our results are not relevant for 
a discussion of the qualitative features of the tail. Therefore, let us write
the lowest order formula for the tail as
$c_q^{\text{as}} = c_q^{\text{rep}} + c_q^{\text{att}}$.
Here, the coefficients of $c_q^{\text{rep}}$ are obtained via a Baxter-like limit,
$\delta\rightarrow0$, K$=\delta/\theta$ fixed. Noting $A\rightarrow0$ in this limit,
we find
\begin{mathletters}\label{eq:cqtail}
\begin{equation}\label{eq:cqtail-a}
c_q^{\text{rep}} = (4\pi/q^2) B_0\cos(q)\;.
\end{equation}
The attraction-induced tail in this approximation reads
\begin{equation}\label{eq:cqtail-b}
c_q^{\text{att}} = (4\pi/q^2) \cdot K 
(2/\delta)\sin(q\delta/2)\,\sin(q(1+\delta/2))\;.
\end{equation}
For the example in Fig.~\ref{fig:fig3}, $(B_0, K) = (4/3, 1/3)$. 
For $q$ below an upper cutoff 
$q_u = \pi/\delta$, the function in the bracket of 
Eq.~(\ref{eq:cqtail-b}) increases almost linearly with $q$. This explains the 
increasing importance of the $c_q^{\text{att}}$ contribution relative to
the $c_q^{\text{rep}}$ one, as is demonstrated in 
Fig.~\ref{fig:fig3}. If $K$ is sufficiently large or if $|B_0|$ is sufficiently small,
one can identify a lower cutoff $q_l = |B_0|/K$ such that $c_q^{\text{att}}$
dominates $c_q^{\text{rep}}$:
\begin{equation}\label{eq:cqtail-c}
c_q^{\text{as}} = (4\pi K/q)\,\sin(q(1+\delta/2))\,,
\quad q_l\ll q\ll q_u\,.
\end{equation}
\end{mathletters}
In the wave-vector interval between $q_l$ and $q_u$, $c_q$ exhibits a power-law
decrease slower than the one of the true large-$q$ tail, which
dominates only for $q\gg q_u$.

\begin{figure}
\centerline{\scalebox{0.75}{\includegraphics{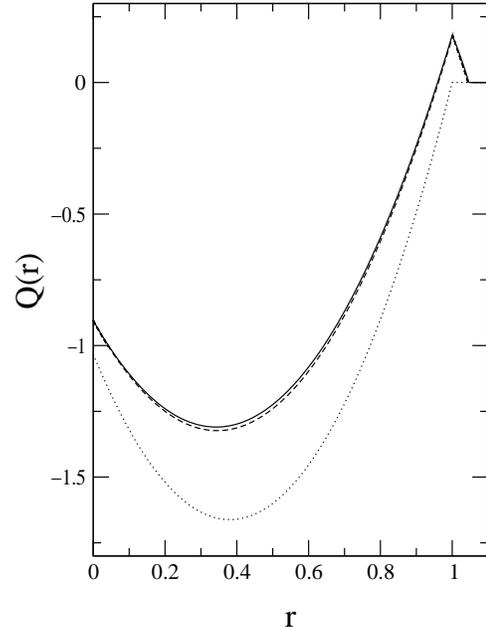}}}
\caption{Factor function $Q(r)$ of the SWS.
The hard-core diameter is chosen as the unit of length.
The dashed line is the PYA result for
$\varphi = 0.5293$, reduced temperature $\theta = 1.10$,
and well width parameter $\delta = 0.0429$. 
The full line is the MSA result for
$\varphi = 0.5258$, $\theta = 0.2332$, $\delta = 0.0465$,
chosen to represent the same physical state of interest in our discussion;
see text for details.
The dotted line shows the result for the HSS at packing fraction $\varphi = 0.516$.
\label{fig:fig4}}
\end{figure}

The PYA and the MSA differ solely by their ansatz for the direct correlation 
function $c(r)$ within the attraction shell $1<r<1+\delta$. Within the PYA,
$c(r)$ depends on $r$ via the $r$-dependence of the pair distribution function,
$c_{\text{PYA}}(r) = \left[1-\exp\left(-u_0/(k_B T)\right)\right] g(r)$,
while the MSA assumes a constant $c_{\text{MSA}} = u_0/(k_B T)$.
In this paper, systems with narrow attraction shells are of interest, 
$\delta\ll 1$. Therefore it is a reasonable approximation to ignore 
the $r$-dependence of $c_{\text{PYA}}$ by writing $g(r)\approx g_d = g(r=1)$.
Thus, for every state where a solution of the PYA exists, there is a solution 
of the MSA, yielding the same structure factor. However, the corresponding 
solution for the MSA has to be evaluated for an effective reduced temperature
$\theta^{\text{MSA}}_{\text{eff}}$. The latter is a smooth function of
$\theta=u_0/k_B T$, $\varphi$ and $\delta$, estimated by
\begin{equation}\label{eq:theta_eff}
\frac{1}{\theta^{\text{MSA}}_{\text{eff}}} \approx
\left[1-\exp\left(\frac{-u_0}{k_B T}\right)\right] g_d\,.
\end{equation}
Consequently, the PYA and the MSA yield the same scenarios for 
the structure factor in the parameter regime of interest in this paper. This 
finding is demonstrated in Fig.~\ref{fig:fig4} for the basic quantity of the
structure factor theory, the factor function $Q(r)$. The result
calculated within the PYA for the parameter triple
\begin{mathletters}\label{eq:A4val}
\begin{equation}\label{eq:A4val-a}
\varphi^*_{\text{PYA}}	= 0.5293,\quad
\theta^*_{\text{PYA}} 	= 1.1000,\quad
\delta^*_{\text{PYA}} 	= 0.0429 
\end{equation}
is very close to the one obtained within the MSA for
\begin{equation}\label{eq:A4val-b}
\varphi^*_{\text{MSA}} 	= 0.5258,\quad
\theta^*_{\text{MSA}} 	= 0.2332,\quad
\delta^*_{\text{MSA}} 	= 0.0465
\,.
\end{equation}\end{mathletters}
It will be shown in Sec.~\ref{sec:phase}, that the two states specified above are of
central importance. The values found for the corresponding densities
$\varphi^{*}$ and well widths $\delta^{*}$ are close to each other.
The difference in the values for the effective attraction strength $1/\theta^*$
is well explained by Eq.~(\ref{eq:theta_eff}). If one inserts $g_d^*\approx7.5$
as obtained from the MSA, one finds $\theta^{\text{MSA}}_{\text{eff}}\approx0.2233$.

All calculations within the MSA are based on the small-$\delta$ expansion for the
factor function, say $Q^{\text{exp}}(r)$, defined by 
Eqs.~(\ref{eq:SWF:Qr}) to (\ref{eq:SWF:abc1}). To control this result,
Eq.~(\ref{eq:WHQ}) and the analog of Eq.~(\ref{eq:PYA}) for the 
MSA closure have also been solved numerically to get the correct MSA factor 
function, say $Q^{\text{MSA}}(r)$. The difference 
$\delta Q(r)=Q^{\text{MSA}}(r)-Q^{\text{exp}}(r)$ is positive and about 
3\% (1\%) for $\delta=0.25,\,(0.15)$ for $r\le 0.5$; and it decreases for $r$ 
increasing above $0.5$.

The dotted line in Fig.~\ref{fig:fig4} exhibits the parabola for $Q(r)$,
Eq.~(\ref{eq:SWF:Qr-a}), for coefficients of the HSS. Introduction of the 
attraction smoothly renormalizes the coefficients $a$, $b$, and $c$, such that
the parabola shifts upwards and becomes flatter. There appears a region of 
positive values for $Q(r)$ near the core surface $r\approx 1$.
These shifts cause the smooth drifts of the Fourier transform for $\hat{Q}(q)$,
which lead to the drifts of $S_q$ discussed in connection with Fig.~\ref{fig:fig2}
and to the appearance of a spinodal, cf.\ Fig.~\ref{fig:fig1}.
The only qualitative new feature, which is caused by the attraction well, is 
the almost straight decrease of $Q(r)$ within the interval $1<r<1+\delta$.
Equation~(\ref{eq:SWF:Qr-c}) yields the slope in leading order as
$Q'(r) = -K/\delta$. In the Baxter limit, this slope diverges. The specified
almost constant part of $Q'(r)$ causes the attraction-tail contribution to the
asymptote of $c_q$, Eq.~(\ref{eq:cqtail-b}). The power law tail, 
Eq.~(\ref{eq:cqtail-c}), is a precursor of the mentioned divergency.

The structure factor or the pair correlation function determine the positions 
of the liquid-gas transition points. However, one faces the known consistency 
problem that different routes for the equation of state yield different results 
for the transition points if approximations for $S_q$ or $g(r)$ are used 
\cite{Hansen86}. We will not discuss these problems in this paper, since 
it is irrelevant for the evolution of glassy dynamics or the glass transition 
whether the fluid is in a stable or metastable thermodynamic state.

\section{The Phase Diagram}
\label{sec:phase}

\subsection{The bifurcation equation}
\label{subsec:phase:bifurcation}

The MCT equations of motion for various dynamical quantities are based on the 
equations for the normalized density correlators $\phi_q(t)=\langle
\rho_{{\vec{q}}}^{*}(t)\rho_{\vec{q}}\rangle/\langle|\rho_{\vec{q}}|^2\rangle$.
For the liquid state these functions approach zero for large time $t$;
density fluctuations which were created at time $t=0$ disappear for 
$t\rightarrow\infty$. The glass state is characterized by a spontaneous arrest 
of these fluctuations, i.e.\ the long-time limits $f_q$ of the correlators do not 
vanish.
The ideal
liquid-glass transition of the MCT is characterized by a discontinuous increase
of $f_q$ from its value zero in the liquid state to the critical Debye-Waller 
factor $f_q^c>0$ of the glass. For colloidal suspensions, $f_q$ can be deduced
from the dynamical-light-scattering results for $\phi_q(t)$.
The experimental findings for the HSS \cite{Megen93b} and for a charge 
stabilized system \cite{Beck99} confirmed the discontinuity for $f_q$ and the 
data for $f_q^c$ agree well with the MCT results.

The $f_q$ obey the equation $f_q/(1-f_q) = {\mathcal F}_q(f)$ \cite{Bengtzelius84}.
Here, the mode-coupling functional ${\mathcal F}_q$ is given by
\begin{mathletters}\label{eq:functional}
\begin{equation}\label{eq:functional-a}
{\mathcal F}_q(f)  = \frac{1}{2} \int{ {{{d}^3k} \over 
(2 \pi )^3} V_{\vec{q},\vec{k}} f_k f_{|\vec{q}-\vec{k}|}}\,.
\end{equation}
The mode-coupling vertices are determined by the structure factor $S_q$,
the direct correlation function $c_q$, and the density $\rho$

\begin{equation}\label{eq:functional-b}
V_{\vec{q},\vec{k}} \equiv  S_q S_k S_{|\vec{q}-\vec{k}|} \rho
\left[ {\vec{q}} \cdot 
\vec{k}\,{c_k} +\vec{q} \cdot  
(\vec{q}-\vec{k})\,{{c_{|\vec{q}-\vec{k}|}} }
 \right]^2/q^4
\;.
\end{equation}
\end{mathletters}
In the following, the wave-vector integrals will be discretized to points on 
a grid of $M$ values, which are equally spaced with step size $h$, starting
at $q_{\text{min}}=h/2$.
Thereby the mode-coupling functional is changed to a second order polynomial
\begin{equation}\label{eq:functional_discrete}
{\mathcal F}_q(f)  = \sum_{k p} V_{q,k p} f_k f_p\,.
\end{equation}
The explicit representation of the coefficients can be found in 
Ref.~\cite{Franosch97}. The $M$ parameters $f_q$ obey the algebraic equations
\begin{equation}\label{eq:bif}
f_q/(1-f_q) = {\mathcal F}_q(f)\,,\quad
q=1, \dots, M\,.
\end{equation}
Besides the long-time limit $f_q$, Eq.~(\ref{eq:bif}) can have further solutions,
say $\tilde{f}_q$, obeying $0\le\tilde{f}_q<1$. The Debye-Waller factor is 
distinguished by the maximum property $f_q\ge\tilde{f}_q,\,q=1, \dots, M$
\cite{LesH}. We used the iteration procedure 
$f_q^{(n+1)}={\mathcal F}_q[f^{(n)}]/(1+{\mathcal F}_q[f^{(n)}])$,
$n=0,\,1,\,\dots$ to determine $f_q$. With increasing $n$ the $f_q^{(n)}$ decrease
monotonically towards $f_q$, if the iteration is started with $f_q^{(0)}=1$ \cite{Goetze95b}.

Two concepts are needed in the following, namely
the maximum eigenvalue $E$ and the exponent parameter $\lambda$ \cite{LesH}.
For the discussion of the implicit equations, Eq.~(\ref{eq:bif}), the Jacobian $J$
is of importance. It is equivalent to $1-C$, where the $M\times M$ matrix $C$
is determined by
\begin{mathletters}\label{eq:matrix}
\begin{equation}\label{eq:matrix-a}
C_{q k} = \frac{\partial {\mathcal F}_q(f)}{\partial f_k}\, 
\left(1-f_k\right)^2\,.
\end{equation}
Also the variation of $C$ with changes of $f$ is needed
\begin{equation}\label{eq:matrix-b}
C_{q,k p} = \frac{1}{2}\,
\frac{\partial^2 {\mathcal F}_q(f)}{\partial f_k\partial f_p}\, 
\left(1-f_k\right)^2\,\left(1-f_p\right)^2\,.
\end{equation}
\end{mathletters}
There is a nondegenerate eigenvalue $E$ of matrix $C$ with the property, that
all other eigenvalues $\tilde{E}$ obey $\tilde{E}<E$.
There holds $E\le 1$, and liquid-glass transition points are determined by the 
condition $E=E_c=1$. It is helpful to follow the drift of $E$ with
changes of control parameters while searching for the transition points.
The left and right eigenvectors of $C$ for the eigenvalue $E$, denoted by 
$\hat{e}$ and $e$ respectively, are uniquely determined by the conditions:
$\hat{e}_q\ge 0,\,e_q\ge 0,\,\sum_q \hat{e}_q e_q =1,\,
\sum_q \hat{e}_q (1-f_q) e_q^2 =1$. They are used to characterize every 
transition point by a single number $\lambda$, defined as 
\begin{equation}\label{eq:lambda}
\lambda = \sum_{q k p} \hat{e}_q^c C^c_{q, k p} e_k^c e_p^c\,.
\end{equation}

The solutions of Eq.~(\ref{eq:bif}), considered as functions of the $M^3$
coefficients $V_{q, k p}$, can exhibit singularities, which are called 
bifurcation points \cite{A92}. The singularities occur if the Jacobian $J$ is
a singular matrix, i.e.\ if the matrix $C$ has eigenvalue unity. The special 
singularities, which are exhibited by the Debye-Waller factors, are called
glass-transition singularities. These are members of the simplest family of
singularities, labeled $A_l,\,l=2,\,3,\,\dots$ \cite{A92}. They are 
topologically equivalent to the bifurcation singularities of the real roots of 
real polynomials of degree $l$. Since the $V_{q, k p}$ are smooth functions 
of the control parameters, in the SWS the Debye-Waller factor $f_q$ exhibits
$A_l$ singularities considered as a function of the variables
$\varphi,\,\theta$, and $\delta$.
The liquid-glass transition is an example for the simplest bifurcation 
singularity $A_2$, called a fold bifurcation. Such transitions 
occur on smooth surfaces in the 3-dimensional parameter space. These surfaces 
can terminate in smooth lines of $A_3$ 
singularities, that are also called cusp bifurcations. The inner points of the
$A_2$-surfaces are characterized by $0<\lambda<1$, and for the endpoints there
holds $\lambda=1$.
The most complicated generic singularity in a three-parameter system
is the meeting of two $A_3$ lines in 
an $A_4$ point. Its position shall be denoted by $\varphi^{*},\,\theta^{*}$, and
$\delta^{*}$. This singularity is also called a swallow-tail bifurcation 
\cite{A92}. The possibility of the described scenarios has been demonstrated
earlier for schematic MCT models, invented with the
mere intention of demonstrating the existence of $A_3$ and $A_4$ points 
\cite{LesH}. This paper is the first demonstration of the existence of an $A_4$
for a microscopic model; the values for the SWS are given in Eq.~(\ref{eq:A4val}).

The numerical work is done with step size $h d = 0.4$. It was checked for
representative cases, that choosing smaller step sizes does not alter the results to
be presented. Choosing $M$ is equivalent to introducing an upper wave-vector
cutoff $q^{*}=(M-1/2)h$ in Eq.~(\ref{eq:functional-a}). The previous 
comprehensive studies for the HSS \cite{Franosch97,Fuchs98}
were done with $M=100$. For sufficiently large $1/\theta$ and sufficiently 
small $\delta$, the direct correlation function $c_q$ develops a large-$q$
tail, discussed in Eq.~(\ref{eq:cqtail}).
This is decisive for fixing the value $q^{*}$ needed to ensure the correct
handling of Eq.~(\ref{eq:functional}).
If the coefficient $K$ in Eq.~(\ref{eq:cqtail-c}) is kept fixed, the cutoff
$q^{*}$ will increase with decreasing well-width parameter
proportional to $1/\delta$. The maximum value for $M$ that can be handled in
the numerical work, defines the lower limit for $1/K$ and $\delta$, which can 
be treated. We used values for $M$ up to $2000$ occasionally, in order
to guarantee the cutoff independence of the results reported in this paper.

\subsection{Results}
\label{subsec:phase:results}

\begin{figure}[t]
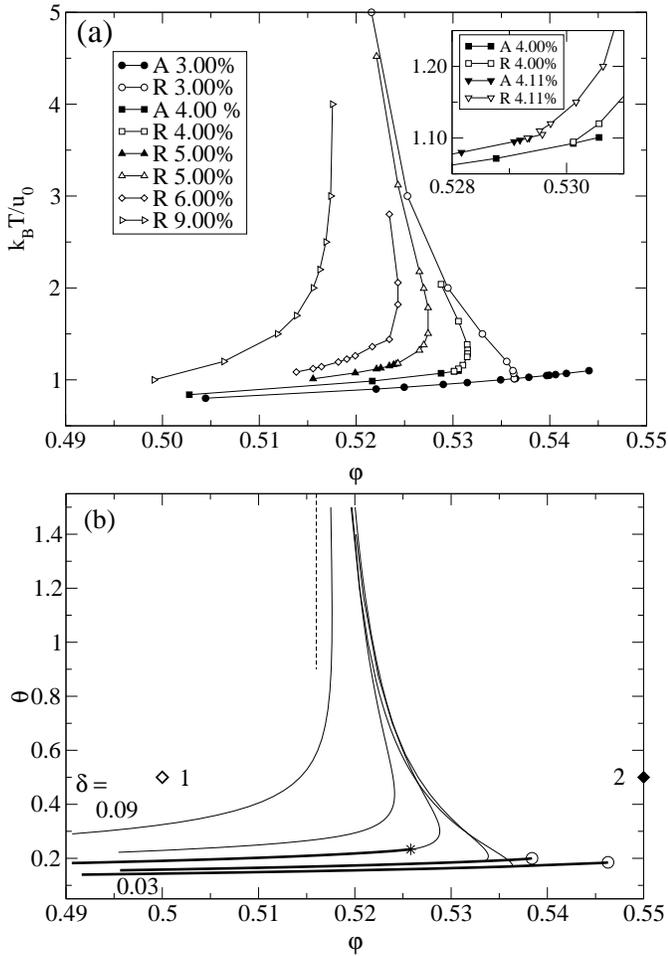

{\centerline{\includegraphics[width=0.49\textwidth]{fig5a.eps}}
\nopagebreak
\centerline{\psfig{figure=fig5b.eps,height=6.25cm,clip=,angle=270.}}}
\caption{The phase diagram of the SWS showing cuts through the control 
parameter space for fixed relative attraction-well width $\delta=\Delta/d$.
The upper part (a) is based on the PYA for the structure factor $S_q$, and the ratio
$\delta/(1+\delta)$ is noted in the legend. The lower part (b) is based on 
the MSA for $S_q$, and the well widths are $\delta=0.09$, $0.06$, $0.0465$, $0.035$,
and $0.03$, subsequently.
The $A_3$ endpoints are marked by open circles and the $A_4$ by an asterisk.
The vertical dashed line marks the transition line $\varphi\approx0.516$
for the hard-sphere system.
For reference, states 1 and 2 from Fig.~\ref{fig:fig1} are included as diamonds.
\label{fig:fig5}}
\end{figure}

\begin{figure}
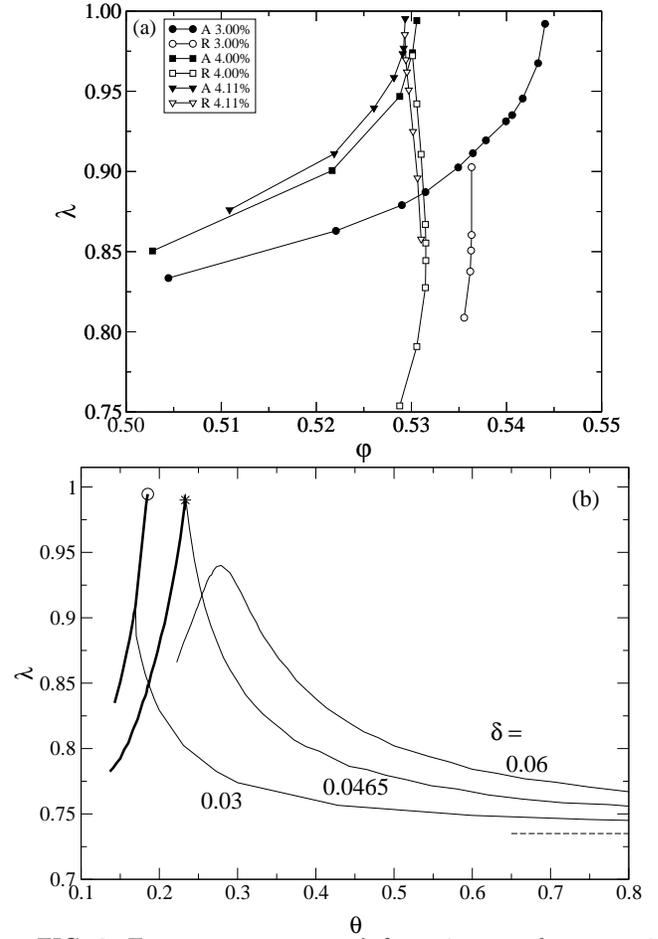

\centerline{\includegraphics[width=0.42\textwidth]{fig6a.eps}}
\nopagebreak
\psfig{figure=fig6b.eps,height=6.25cm,clip=,angle=270.}
\caption{Exponent parameter $\lambda$ for points on three transition lines. The upper part
(a) was calculated within the PYA for the ratios $\delta/(1+\delta)$ noted in the legend.
Part (b) shows the results for the MSA, where the
dashed line indicates the value $\lambda=0.735$ of the HSS.
\label{fig:fig6}}
\end{figure}

The phase diagram for the SWS is shown in Fig.~\ref{fig:fig5} for several
constant-$\delta$ cuts through the three dimensional control-parameter space. 
The results based on the PYA and the MSA are qualitatively the same. 
Let us first consider the three states 1, 2, and 3 from Fig.~\ref{fig:fig1}
for $\delta=0.06$. Within the MSA, state~1 refers to the liquid phase,
cf.\ Fig.~\ref{fig:fig5}b. Increasing $\varphi$ to the state~2 increases the 
height of the first sharp diffraction peak of $S_q$, located near $q_0= 7$,
cf.\ Fig.~\ref{fig:fig2}. Thus the compressibility for fluctuations in the shell
$q\approx q_0$ increases. This leads to arrest in a glass state, as known from the
HSS.
If one cools state~1 at fixed $\varphi=0.50$ down to state~3,
$S_{q_0}$ decreases, as was explained in connection with Fig.~\ref{fig:fig2}.
This effect stabilizes the liquid, but it is overcompensated
by the increase of $S_q$ in the small-wave-vector region, $q < 6.1$, and,
more important, in the large-$q$ region, $q > 7.4$.
As a result of this compressibility increase on the wings of the structure 
factor peak, the liquid freezes to a glass upon cooling, cf.\ Fig.~\ref{fig:fig5}.
For large temperature, $S_q$ depends only weakly on $T$; the terms
proportional to $K=\delta\cdot u_0/(k_B T)$ in 
Eqs.~(\ref{eq:SWF:Qr}) to (\ref{eq:SWF:abc1}) cause only small modifications 
of the coefficients determining the factor function $Q(r)$.
This explains, why the transition lines are almost vertical in 
Fig.~\ref{fig:fig5}a for $k_B T/u_0 > 3$ and in Fig.~\ref{fig:fig5}b for
$\theta > 1$. The peak wings are not very sensitive to density changes by a
few percent for $\varphi<0.51$. This explains, why the transition lines in
Fig.~\ref{fig:fig5} are rather flat there.
The two pieces of the transition line join smoothly and $\lambda$ remains below unity
for $\delta=0.06$,
as is shown in Fig.~\ref{fig:fig6}b. Thus, the described curve represents a cut
through a smooth surface of $A_2$ bifurcations.

The mentioned high-temperature pieces of the transition surface are 
located at packing fractions $\varphi_c$, which exceed the value for the HSS, 
$\varphi_c^{\text{HSS}} \approx 0.516$. This means, that the attraction forces have
stabilized the liquid phase. This effect is smaller for larger $T$ and therefore the
$\varphi_c$-versus-$T_c$ curve decreases. There is the possibility of glass
melting due to cooling, if the decrease of $S_{q_0}$ is not overcompensated
by the increase of the structure-factor-peak wings. The attraction causes
bonding, in the sense that the 
average separation of two particles is smaller than expected for a HSS.
Therefore the average size of the holes increases and this favors the 
long-distance motion characteristic for a liquid. Consider states with
$\varphi=0.52$ and $\delta=0.06$. For the MSA results one notices from
Fig.~\ref{fig:fig5}b, that the system is in a glass for $\theta = 0.10$
and it melts upon heating if $\theta$ approaches $\theta_m^{-} \approx 0.30$.
This transition occurs due to the temperature drift of the coupling to modes
with wave vectors in the wings of the $S_q$ peak. The system remains in the 
liquid upon further heating until it reenters the glass at
$\theta_m^{+} \approx 1.41$.
This freezing is caused by the drift of the coupling to modes with $q\approx 
q_0$. The described reentry phenomenon \cite{Fabbian99,Bergenholtz99} is
a manifestation of two mechanisms for localization due to the cage effect
in the high density SWS to be explained below.

The preceding two paragraphs can be summarized as follows. There is a subtle
interplay of excluded-volume and bonding effects which determine the 
variations of the pair correlation function $g(r)$ on the length scale of the
particle diameter $d$. This is reflected in the properties of the structure
factor $S_q$ for wave numbers $q$ at and around the peak position $q_0$; the
relevant $q$ range is the one exhibited in Fig.~\ref{fig:fig2}. Fluctuations
with longer wave length are of no qualitative importance for those parameter
points studied in this paper. This conclusion was corroborated by dropping
all contributions to the mode-coupling functional, 
Eq.~(\ref{eq:functional_discrete}), where $k$ or $p$ are smaller than $4$;
there was no significant change of the phase diagram calculated with the MSA 
structure compared to what is shown in Fig.~\ref{fig:fig5}b.
Similarly, a cutoff $q_{+}=20$ was introduced such that all contributions to 
the mode-coupling functional with $q>q_{+}$ are dropped. The MSA results shown
in Fig.~\ref{fig:fig5}b for $\delta=0.09$ and $\delta=0.06$ did not change, 
nor was there a noticeable change for the other curves for $\theta > 0.6$.
We conclude, that the two specified
sources for correlations on intermediate length scale explain the phase 
transition points, which are marked by open symbols in Fig.~\ref{fig:fig5}a
or by the corresponding light lines in Fig.~\ref{fig:fig5}b.

To substantiate the previous conclusion, we have constructed a further phase
diagram based on the MSA structure factor, using
the above specified cutoff $q_{+}$. As mentioned, the curves for $\delta=0.09$
and $\delta=0.06$ were reproduced up to minor deviations. Upon decreasing $\delta$
further, curves emerge, which continue the trend of the
two ones for larger $\delta$. The limit $\delta \rightarrow 0$ can be 
carried out; no new features appear. This ad-hoc MCT model yields a smooth liquid-glass
transition surface of $A_2$-bifurcation points with $\lambda<1$.
Obviously, the true phase diagram of the
SWS shown in Fig.~\ref{fig:fig5} is quite different. There are the transition points 
marked by filled symbols in Fig.~\ref{fig:fig5}a or by the corresponding heavy 
lines in Fig.~\ref{fig:fig5}b. These define transition lines, which do not
join smoothly the previously discussed lines. Rather they cross the former
lines. Thus, for sufficiently small $\delta$ and sufficiently large attraction
strength $u_0/(k_B T)$, there is a new glass formation mechanism, dominated by 
density fluctuations with large wave number $q\ge q_{+}$.
These are due to spatial correlations on the length scale of the 
attraction-well width $\Delta$.

For a discussion of the identified new pieces of the transition surface, the
mode-coupling coefficients in Eqs.~(\ref{eq:functional-a}) and 
(\ref{eq:functional-b}) can be simplified. As explained in connection with 
Fig.~\ref{fig:fig3}, one can write for $k=|\vec{k}|\ge q_{+}$ and 
$p = |\vec{q}-\vec{k}|\ge q_{+}$ the leading asymptotic expression for the 
structure factors, $S_k = S_q = 1$, and for the direct correlations function
$c_q = c_q^{\text{as}}$, Eq.~(\ref{eq:cqtail-b}). Thus, the dominant part of 
the mode-coupling functional depends explicitly on the control parameters by 
the prefactor $\varrho K^2 = \varrho\left(\delta u_0/(k_B T)\right)^2$ and
otherwise only on $\delta$ via the large-wave-vector cutoff $q_u$. A further
density dependence is due to fluctuations with $q<q_{+}$ only. This explains
why the low-$T$-transition lines in Fig.~\ref{fig:fig5} are so flat. $K^2$
decreases proportional to $\delta^2$, and this effect is not overcompensated by the increase of 
$q_u$. As a result, the horizontal transition lines decrease with decreasing $\delta$.
For low packing fractions, such a trend can be shown explicitely by an analytic
calculation \cite{Bergenholtz99,Bergenholtz00}.
Along this transition line, $\lambda$ increases with increasing 
$\varphi$ until it approaches unity signalizing an endpoint at some 
$\varphi_c^{\circ}(\delta),\,\theta_c^{\circ}(\delta)$. 
This is demonstrated in Fig.~\ref{fig:fig6}a for three values of $\delta$ and
in Fig.~\ref{fig:fig6}b for $\delta=0.03$. The line stops the previously discussed
line in some crossing point. Between the crossing point and the endpoint, there occur
glass-glass transitions. Upon increasing $\delta$, the length of the glass-glass transition
line shrinks, until it vanishes for some $\delta^{*}$ at some $\varphi^{*}=
\varphi_c^{\circ}(\delta^{*})$ and $\theta^{*}=\theta_c^{\circ}(\delta^{*})$;
and this is the $A_4$ singularity whose coordinates are listed in 
Eq.~(\ref{eq:A4val}).

\begin{figure}
\centerline{\psfig{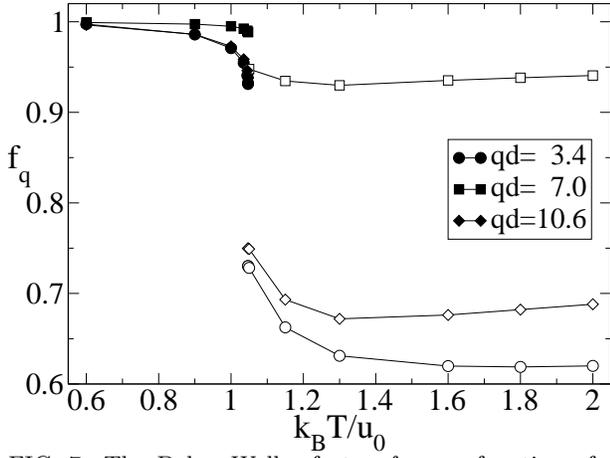}}
\caption{The Debye-Waller factor $f_q$ as a function of reduced temperature for
fixed packing fraction $\varphi=0.539672$ and fixed $\delta/(1+\delta)=0.03$.
The wave vector $qd = 7.0$ is close to the structure factor peak position.
The calculations are based on
the PYA for $S_q$. The path through the parameter space deals with a 
glass-glass transition occuring at $\theta_c=1.0471$, compare
Fig.~\ref{fig:fig5}a.
\label{fig:fig7}}
\end{figure}

Every pair of glass states can be connected by a curve in parameter space such
that, upon shifting the control parameters $\varphi$, $\theta$, and $\delta$
along this curve, the glass properties change smoothly. Thus one cannot 
discriminate precisely between repulsion- and attraction-dominated glass 
states. However, upon crossing the glass-glass-transition surface, there occurs
a discontinuous change of the glass-state properties. These discontinuities
can be used to differentiate quantitatively between the two types of glasses.
From a mathematical point of view, the situation is analogous to the termination
of the liquid-gas-transition line at the critical point.
Some results shall be presented from the theory based on the PYA
for $S_q$. Let us consider states for $\varphi \approx 0.54$ 
for the smallest $\delta$ used in Fig.~\ref{fig:fig5}a. Figure~\ref{fig:fig7}
exhibits as filled symbols the decrease of the Debye-Waller factor for three
representative wave numbers, if the attraction-dominated glass is heated
towards the transition temperature $T_c$, $k_B T_c/u_0 = 1.0471 = \theta_c$.
The $f_q$ decrease towards the critical value $f_q^c$ according to the 
asymptotic square root law $f_q-f_q^c\propto h_q \sqrt{\theta_c-\theta}$, 
which is the signature of the fold bifurcation \cite{LesH}. Upon
crossing the line into the repulsion-dominated glass, $f_q$ drops and keeps
on decreasing upon further heating up to $\theta\approx 1.3$, as shown by the
open symbols. Notice, that $f_q$ does not exhibit any singularity for $\theta$
decreasing towards $\theta_c$. The remarkable variation of $f_q$ for $\theta$
near but above $\theta_c$ is a precursor phenomenon of the nearby $A_3$ 
singularity. The $f_q$ for $\theta>\theta_c$ is smaller than the Debye-Waller
factor for the HSS at the same packing fraction. Hence the attraction has
softened the glass. This effect has to disappear for very large $T$, and this
explains, why $f_q$ increases again, reflecting a glass
stiffening upon heating. The described effects for $\theta>\theta_c$ are the
counterparts to what was discussed above in connection with the reentry phenomenon.

\begin{figure}
\centerline{\psfig{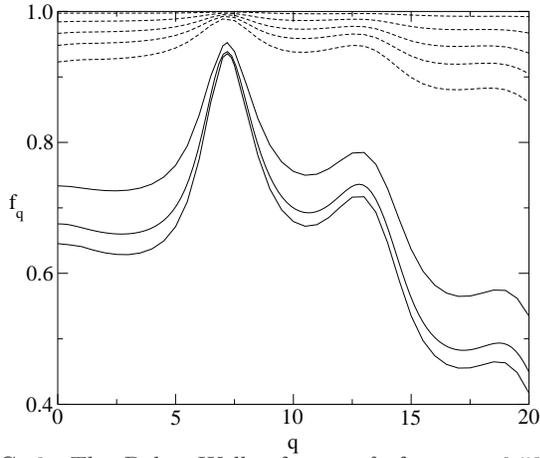}}
\caption{The Debye-Waller factors $f_q$ for $\varphi=0.539672$,
$\delta/(1+\delta)=0.03$. The temperature increases from top to bottom as 
$\theta=k_B T /u_0=0.600,\,0.900,\,1.000,\,1.035,\,1.0471$ (dashed lines) and
$\theta=1.047,\,1.150,\,1.300$ (full lines). The calculations are based on
the PYA for $S_q$.
\label{fig:fig8}}
\end{figure}

The wave-vector dependence of $f_q$ changes qualitatively upon crossing the
glass-glass-transition line as is shown in Fig.~\ref{fig:fig8}. Let us focus
in this paragraph on the wave-vector regime at and above the structure factor
peak position $q \gtrsim q_0 \approx 7$. Here, $f_q$ oscillates with 
wave-vector scale $2\pi/d$ around the M\"o{\ss}bauer-Lamb factor $f_q^s$. The latter is
the analog to $f_q$, constructed for a tagged particle with position
vector $\vec{r}_s(t)$ via its density-correlator 
$\phi_q^s(t) = \langle \rho^s_{\vec{q}}(t)^{*} \rho^s_{\vec{q}}\rangle$;
$\rho^s_{\vec{q}}(t) = \exp\left(i \vec{q}\vec{r}_s(t)\right)$. This quantity,
in particular its long-time limit $f_q^s$, can also be measured 
\cite{Megen98}. One finds $f_q^s = 1-q^2 r_l^2 + {\mathcal O}(q^4)$, where
$r_l$ is the localization length of the particle: $r_l^2 = \lim_{t\rightarrow
\infty} \langle|\vec{r}_s(t)-\vec{r}_s(0)|^2\rangle$ \cite{LesH}.
In the Gaussian approximation, one can write $f_q^s=\exp\left(-q^2r_l^2\right)$
so that the half-width-wave vector $q_l$, defined by $f_{q_l} = 0.5$, can be
used to estimate $r_l \approx 1/q_l$. For the Debye-Waller factors on the 
high-temperature side of the transition line, which are shown as full lines in
Fig.~\ref{fig:fig8}, one estimates $q_l\approx 20$. The localization length
$r_l=0.05$ is about the same size as expected for a particle rattling
between the hard walls of its cage in a HSS. The $f_q$ and $r_l$ are close to those
of a HSS \cite{LesH,Fuchs98}. However, on the low-temperature side of the
transition, the $f_q$
vary by less than 10\% if $q$ increases up to $2 q_0$, as shown by the dashed
lines in Fig.~\ref{fig:fig8}. The wave number $q_l$ is much larger than expected
from the free volume in the cage. At the transition, $q_l\approx 44$, i.e. the
localization length is decreased discontinuously by a factor of about 2.3. For
$\theta=0.9$ and $0.6$ the localization length is about $0.01$ and 
$0.006$, respectively. This shows that the localization length is of order 
$\delta$. The particle is bound to the wall of the cage and localization is 
determined entirely by the particle attraction.
This localization mechanism via bond formation is operative at low
packing fractions also, and it has been studied within MCT in this regime.
There, bond formation has been argued to be of importance for colloidal gelation
\cite{Bergenholtz99,Bergenholtz00,Bergenholtz99b}. 

\begin{figure}
\centerline{\includegraphics[width=0.45\textwidth]{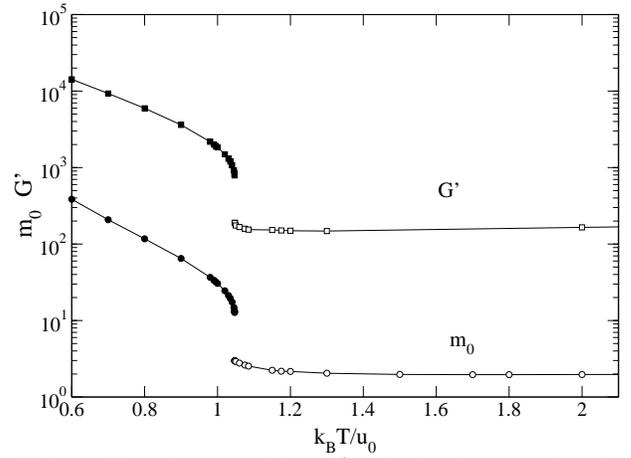}}
\caption{The shear modulus $G'$ and the dimensionless longitudinal elastic modulus
$m_0$ as a function of the reduced temperature. The 
result is based on the PYA for $S_q$, and the parameters of state deal with the 
same path through the glass-glass transition as discussed in Figs.~\ref{fig:fig7} 
and \ref{fig:fig8}.
\label{fig:fig9}}
\end{figure}

There is no strong dependence of $f_q$ on wave numbers $q<4$. The 
low-temperature glass is distinguished from the high-temperature one by the 
fact that the zero-wave-number limit of the Debye-Waller factor, $f_0$, is
larger for the former than for the latter. Therefore, upon crossing the 
glass-glass-transition surface by cooling, the peak for $q\approx q_0$ of the
$f_q$-versus-$q$ diagram disappears, Fig.~\ref{fig:fig8}. The number $f_0$ is
related to the longitudinal elastic modulus of the system. This consists of a 
part expected for the ergodic liquid and a part $m_0$ reflecting the incomplete
relaxation of the non-ergodic glass \cite{LesH}. The latter is given by the 
zero-wave-number limit of the mode-coupling functional, 
Eq.~(\ref{eq:functional_discrete}), $m_0 = {\mathcal F}_0(f)$. The $q=0$ limit
can be carried out easily in Eq.~(\ref{eq:functional-b}), so that one derives from
Eq.~(\ref{eq:functional-a}) a formula \cite{Bengtzelius84},
\begin{equation}\label{eq:mL}
m_0 = \int_0^\infty {d}k\, v^L(k) f^2_k
\,.
\end{equation}
After discretization, one can substitute the results for $f_k$ to get $m_0$, 
and via Eq.~(\ref{eq:bif}) one has $f_0 = m_0/\left(1+m_0\right)$. For the 
glass with $\theta>\theta_c$, the integral in Eq.~(\ref{eq:mL}) is dominated
by fluctuations on the intermediate wave-vector scale, $k\approx q_0$. As
known from the HSS, $m_0$ is of order unity and thus $f_0$ is about $0.5$.
However, for $\theta\le \theta_c$, the integral is dominated by large-$k$
fluctuations. This enhances $m_0$. For  $\theta =\theta_c$, one finds 
$m_0\approx 12.7$ and this increase of the modulus explains the increase of $f_0$
to about $0.927$, exhibited by the lowest dashed line in Fig.~\ref{fig:fig8}.
Decreasing the temperature to $k_B T/u_0 = 0.6$, leads to $m_0\approx 385$ and
this explains the large value $f_0\approx 0.997$, exhibited by the uppermost
dashed curve in Fig.~\ref{fig:fig8}. For the shear modulus $G'$, a
formula like Eq.~(\ref{eq:mL}) holds, where the expression for $v^T$ is similar to
that for $v^L$ \cite{Bengtzelius84}. Therefore, the shear modulus also exhibits the 
specified strong enhancement due to the attraction wells.
Figure~\ref{fig:fig9} shows that the dramatic change of the moduli is the most 
relevant feature to be observed upon crossing the glass-glass-transition 
surface. The strong short-ranged attraction causes bond formation, and this 
increases the rigidity of the glass with respect to compressions or shearing
considerably relative to that of a glass at a similar density, where the 
structural arrest is dominated by mere hard-sphere repulsion.

\section{Structural Relaxation}\label{sec:dynamics}

\subsection{Some general MCT equations}\label{subsec:dynamics:general}

The MCT equations of motion are based on the exact expression of the density correlator
$\phi_q(t)$ in terms of a fluctuating-force correlator \cite{Hansen86,Balucani94},
which in turn is split into a part treating normal liquid effects, and a relaxation
kernel $m_q(t)$ describing the cage-effect contribution. This kernel is approximated
by the mode-coupling functional, discussed above in connection
with Eqs.~(\ref{eq:functional}) and (\ref{eq:functional_discrete}):
$m_q(t)={\mathcal F}_q\left[\phi(t)\right]$, \cite{Bengtzelius84,LesH}. This paper
will be restricted to the simplest approximation for the normal liquid effects,
i.e.\ the colloid will be treated as a system of Brownian particles, so that only the
instantaneous correlations as given by the structure factor $S_q$ are incorporated.
As a result one obtains \cite{Franosch97}:
\begin{equation}\label{eq:dynamics}
\tau_q\partial_t\phi_q(t)+\phi_q(t)+\int_0^t m_q(t-t')\partial_{t'}\phi_q(t')\,dt'=0\;.
\end{equation}
This equation implies the short-time asymptote $\phi_q(t)=1-(t/\tau_q)+
{\mathcal O}(t^2)$. For the time scale, one finds $\tau_q=S_q/(D_0q^2)$, where
$D_0$ denotes the single-particle diffusion coefficient. $D_0$ reflects the property
of the solvent, and it fixes the time scale for the transient motion. The unit of
time shall be chosen such that $1/D_0=160$ to ease comparisons with the results
for the HSS from preceding work \cite{Franosch97,Fuchs98}.

Two comments on the implications of Eq.~(\ref{eq:dynamics}) might be appropriate. First,
the solutions are completely monotone functions, i.e.\ there is 
a rate density $\rho_q(\gamma)\ge 0$, normalized to 
$\int_0^\infty \,\rho_q(\gamma)\,d\gamma = 1$, such that
\begin{equation}\label{eq:rate_distribution}
\phi_q(t)=\int_0^\infty e^{-\gamma\,t}\,\rho_q(\gamma)\,d\gamma\;.
\end{equation}
Thus, the MCT approximations maintain a fundamental property of colloidal dynamics:
auto-correlation functions can be written as superpositions of Debye-relaxation
functions \cite{Goetze95b}. Second, outside the transient, the solutions can be
written as $\phi_q(t)=F_q(t/t_0)$. Here, $F_q$ is given by the mode-coupling functional
${\mathcal F}_q$, i.e.\ by the equilibrium structure factor $S_q$. The transient
dynamics, no matter how complicated, enters via the single time scale $t_0$ only
\cite{Goetze92,Franosch98,Fuchs99b}. The following results for the long-time
dynamics are thus not influenced by the simplified treatment of the short-time
dynamics in Eq.~(\ref{eq:dynamics}), except up to a change of the overall time scale
$t_0$. It is known that the short-time dynamics in colloids is influenced by
hydrodynamic interactions \cite{Pusey91}. Unfortunately, it is not known how to
incorporate these interactions in a theory for high-density colloids. But we consider
it plausible, that the hydrodynamic interactions merely renormalize the transient
dynamics \cite{Fuchs99c}, thereby being irrelevant for the structural-relaxation
effects.

For control parameters approaching a glass-transition singularity, there appears
an increasingly larger dynamical window, where the solutions are
arbitrarily close to the critical Debye-Waller factor $f_q^c$. Therefore, one can
solve the MCT equations of motion by an asymptotic expansion, using
$\delta\phi_q(t)=\phi_q(t)-f_q^c$ as a small parameter. The result can be expressed
in the form
\begin{equation}\label{eq:dyn_asymptotics}
\phi_q(t)-f_q^c=h_q G(t)+h_q^{(1)}G^{(1)}(t)+\cdots\;.
\end{equation}
Most of the known universal results for the MCT bifurcation dynamics are based on the
understanding of the leading-order contribution $G(t)$. The next-to-leading-order
term $G^{(1)}(t)$ allows to discuss the range of validity of the leading-order formulas.
A comprehensive demonstration of the cited results for
the HSS can be found in Refs.\ \cite{Franosch97} and \cite{Fuchs98}.

Equation~(\ref{eq:dyn_asymptotics}) reduces to a particular transparent form for the
critical dynamics, i.e.\ for control parameters placed on the glass-transition
points. For the fold bifurcation, one finds a power-law decay:
\begin{mathletters}
\begin{equation}\label{eq:critical_a2}
G(t)=(t_0/t)^a,\quad G^{(1)}(t)=(t_0/t)^{2a}\;;\quad A_2\;.
\end{equation}
The critical exponent $a$ is given by the exponent parameter $\lambda$ via
$\Gamma(1-a)^2/\Gamma(1-2a)=\lambda$, $0<a\le0.5$. The endpoints of the
$A_2$-bifurcation surfaces are characterized by exponent $a$ approaching zero.
For the cusp bifurcation, there holds \cite{Goetze89d}:
\begin{eqnarray}\label{eq:critical_a3}
G(t)&\propto&1/\ln(t/t_0)^2, \nonumber\\
G^{(1)}(t)&\propto&\ln(\ln(t/t_0))/\ln(t/t_0)^3\;;\quad A_3\;.
\end{eqnarray}
For the swallow-tail bifurcation, one has \cite{Goetze89d,Sperl00}:
\begin{eqnarray}\label{eq:critical_a4}
G(t)&\propto&1/\ln(t/t_0), \nonumber\\
G^{(1)}(t)&\propto&\ln(\ln(t/t_0))/\ln(t/t_0)^2\;;\quad A_4\;.
\end{eqnarray}
\end{mathletters}
The dependence of the leading-order contribution $G(t)$ on the control parameters
is well understood, but shall not be considered in this paper.

\subsection{Results}\label{subsec:dynamics:results}

Let us first estimate the dynamical window relevant for the discussion.
For the HSS, the correlators $\phi_q(t)$ decay from $1.00$ to
$0.95$ for times increasing up to about $t=1$ for representative wave vectors.
In this sense, $t=1$ is the scale for the transient dynamics. After a crossover window
of about one or two decades, the leading-order asymptotic law
$\phi_q(t)=f_q^c+h_q G(t)$ becomes valid at about $t=10^2$.
This value may be an order of magnitude smaller or larger, depending on the
wave number $q$ \cite{Franosch97}. The same is true for the data obtained by
van~Megen \emph{et~al.} for hard-sphere colloids, provided one identifies the time
unit $t=1$ with $1\,\text{msec}$ \cite{Megen95}. The correlators have been measured
up to $10^6\,\text{msec}$, and thus the so far explored windows extend up to
$10^6$ in the units used here. This limit might shift up in future work, using
different experimental setups.

\begin{figure}
\centerline{\includegraphics[width=.32\textwidth,angle=270]{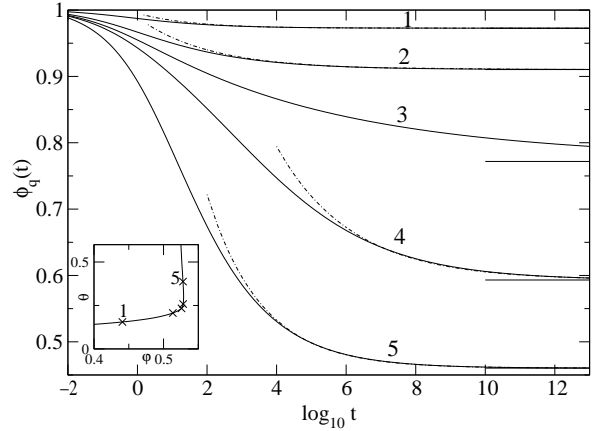}}
\caption{The full lines are the critical correlators $\phi_q^c(t)$ for the
wave vector $q=4.2$ calculated with the MSA-structure factor for the critical
attraction well width $\delta^{*}=0.0465$. The states are located on the
transition line as shown in the inset and refer to critical Debye-Waller
factors $f_q^c = 0.973$, $0.910$, $0.772$, $0.593$, $0.460$ (from top to bottom as
indicated by horizontal straight lines).
For states 1 and 5, $\lambda = 0.80$ corresponding to a critical exponent $a = 0.279$;
state 2 (4) refers to $\lambda = 0.895$ ($0.908$) corresponding to $a = 0.210$ ($0.202$).
The dash-dotted lines show the asymptotes $f_q^c + A_q t^{-a}$.
State 3 is at the $A_4$ singularity, given by Eq.~(\ref{eq:A4val-b}).
\label{fig:fig10}}
\end{figure}

Figure~\ref{fig:fig10} exhibits the critical correlators $\phi_q^c(t)$ for $q=4.2$
for five states on the transition line $\delta=0.0465$ through the $A_4$ singularity.
The states 1 and 5 refer to an exponent parameter $\lambda\approx0.80$ on the side of the
attraction-dominated and repulsion-dominated glass, respectively. For times of the
order of $10^3$ and larger, the leading-order formula, $\phi_q(t)-f_q^c\propto
(t_0/t)^a$, $a\approx0.28$, describes the results. Thus, the scenario is similar to the
one known from the HSS, and this is also true for other states on the line with
$\lambda<0.80$. However, if one considers states closer to the $A_4$ point, the
onset of the critical power law gets shifted to larger times. This is demonstrated
for the two states 2 and 4, which deal with $\lambda\approx0.90$. For the state 4, the
$t^{-a}$ law with $a\approx0.20$ is valid only for $t>10^6$. This trend continues, if one
moves even closer to the $A_4$ point, whereby $\lambda$ increases even further; compare
Fig.~\ref{fig:fig6}. At the $A_4$ singularity, the correlator decays from $1.00$ to
$f_q^c\approx0.77$. This decay is stretched so enormously, that even for $t=10^{12}$
it only reaches the value $\phi_q(t)\approx0.80$, as is shown by curve 3 in
Fig.~\ref{fig:fig10}. One can describe the critical
correlator with Eqs.~(\ref{eq:dyn_asymptotics}) and (\ref{eq:critical_a4}) for the
window $10^{15}<t<10^{25}$, using $h_q$ and $h_q^{(1)}$ as fit parameters. But this
fit does not describe the correlator for $t<10^{10}$. Thus one concludes, that the
critical correlator of the $A_4$ for the SWS cannot be described by the asymptotic
Eq.~(\ref{eq:critical_a4}) within accessible dynamical windows. Nor can the critical
power-law decay of the $A_2$ singularity be measured, if $\lambda$ exceeds a certain
value, say $0.9$.
Thus, there is a part of the transition lines near the $A_4$ point, characterized by
$\lambda\ge0.9$, where the correlators exhibit structural relaxation patterns towards
the plateau values $f_q^c$ that are stretched up to $t=10^6$ or larger. The known
asymptotic formulas cannot be used to describe the MCT solutions within
this regime.

\begin{figure}
\begin{center}%
\includegraphics[width=.32\textwidth,angle=270]{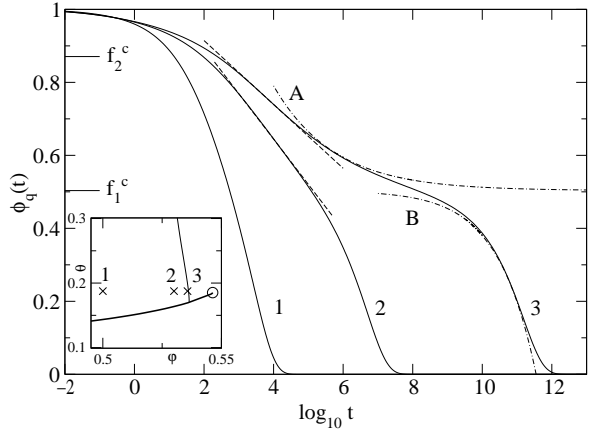}%
\end{center}
\caption{Correlators $\Phi_q(t)$ for $q = 4.2$ calculated for the
MSA-structure factor of a SWS with attraction-well width $\delta = 0.03$ and
the reduced temperature $\theta = 0.1875$ for the three packing fractions 
$\varphi_1 = 0.5000,\,\varphi_2 = 0.5300,\,\varphi_3 = 0.5357$ (full lines).
The inset shows the relevant section from the phase diagram of
Fig.~\ref{fig:fig5}(b).
The dashed-dotted lines with labels A and B exhibit
the critical law $f_q^{(1)\,c} + A_q/t^{0.250}$ and
the von-Schweidler law $f_q^{(1)\,c} - B_q t^{0.396}$, respectively.
The straight dashed lines exhibit logarithmic decay laws,
Eq.~(\ref{eq:logarithm}), see text.
\label{fig:fig11}}
\end{figure}

The liquid dynamics on the small-$\delta$ side of the $A_4$ point is particularly
subtle, since there is an $A_3$ singularity in addition to the line-crossing point.
Figure~\ref{fig:fig11} exhibits as an example such a
situation for $\delta=0.03$. Parameters on a straight line, $\theta=0.1875$,
which is slightly above the $A_3$ point, are considered. The transition to a
repulsion-dominated glass state then occurs at $\varphi_c=0.5360$. At the
transition point, the critical Debye-Waller factor is $f_q^{(1)c}\approx0.50$, and the
exponent parameter is given by $\lambda=0.847$, implying a critical exponent $a=0.250$.
Curve 3 was calculated for such small distance from the transition point,
$-\varepsilon=(\varphi_c-\varphi)/\varphi_c=7.9\cdot 10^{-4}$,
that $\phi_q(t)$ decays to zero just within the dynamical window displayed
in the figure. The dash-dotted line with label A presents the leading-order critical
law for the $A_2$ singularity, Eqs.~(\ref{eq:dyn_asymptotics}) and
(\ref{eq:critical_a2}), with the time scale $t_0$ fitted to the decay at long times
for $\varphi=\varphi_c$. One observes the same phenomenon as explained above in
connection with Fig.~\ref{fig:fig10}: since $\lambda$ is rather large, the asymptotic
law describes the data only for rather large times, $t>10^{5.5}$. After falling below
the plateau value $f_q^{(1)c}$, the correlator decays towards zero. This is the
$\alpha$ process, and it  starts with the von-Schweidler asymptote
$f_q^{(1)c}-h_q(t/\tau)^b$,
which is shown by the dash-dotted line with label B. The exponent $b=0.396$ obeys a
similar relation as the critical exponent, $\Gamma(1+b)^2/\Gamma(1+2b)=\lambda$.
Thus the structural relaxation
connected with the liquid-glass transition follows the known scenario, except that
the familiar $A_2$ patterns can be observed only for times far out the transient
regime, $t>10^{5.5}$.

Figure~\ref{fig:fig11} exhibits a large dynamical window, $10^2<t<10^{5.5}$, where
the structural relaxation does not follow the asymptotic laws for a fold bifurcation.
Instead, the dashed straight line demonstrates that the correlator labeled 3 follows a
logarithmic decay law,
\begin{equation}\label{eq:logarithm}
\phi_q(t)=f_q^{(2)c}-C_q\ln t\;,
\end{equation}
for the major part of the mentioned window, $10^2\le t\le10^{4.5}$. Here,
$f_q^{(2)c}\approx0.87$ is the Debye-Waller factor at the $A_3$ singularity.
There is a line through every $A_3$ singularity, which is transversal to the transition
line ending at the $A_3$, such that Eq.~(\ref{eq:logarithm}) is a leading-order solution
for the MCT equations of motion on a certain intermediate time window. The length of the
window expands and the prefactor $C_q$ in front of the $\ln t$ decreases, if one moves
closer towards the $A_3$ point \cite{Goetze88b}. These results explain the appearance of
the $\ln t$ part and the change of its prefactor,
if one compares curve 2 with curve 3 in Fig.~\ref{fig:fig11}. One
concludes that it is the bifurcation dynamics of the $A_3$ singularity which prevents
the evolution of the $t^{-a}$ law for the fold bifurcation. Similarly,
the $\alpha$ process for curve 2 does not start with von Schweidler's law. Therefore,
contrary to what one observes for the dynamics of the HSS for comparable large times
\cite{Franosch97}, the $\alpha$ processes for curves 2 and 3 do not obey the
superposition principle. Close to the $A_3$ point, the dynamics outside the transient
and preceding the onset of the $\ln t$-decay law follows the critical law for the
$A_3$, as given by Eqs.~(\ref{eq:dyn_asymptotics}) and (\ref{eq:critical_a3}). But for
the shown curves, the situation is similar as explained comprehensively for the critical
decay for the HSS \cite{Franosch97}. The plateau $f_q^{(2)c}$ is so high, that there
is only a small variation remaining for the $1/(\ln t)^2$ law to manifest itself.
The correction
terms for the cited leading-order and next-to-leading-order contributions are so large,
that one has to consider states much closer to the $A_3$ to see the result of
Eq.~(\ref{eq:critical_a3}).

For state 1, the $A_2$, $A_3$, and $A_4$ singularities are so far away, that none of
the cited asymptotic laws is clearly developed. On the other hand, they are close
enough to cause a considerable relaxation stretching. The correlator $\phi_q(t)$
needs a dynamical window of three orders of magnitude to complete the $80\%$ of its
decay from $0.9$ to $0.1$, as is shown in Fig.~\ref{fig:fig11}.

In Fig.~\ref{fig:fig12}a, a set of correlators $\phi_q(t)$ for five representative
wave numbers is shown. The state refers to the liquid close to the $A_4$ point. For
$q=24.2$, the window for the logarithmic decay extends from $t=10^3$ to $t=10^{10}$.
For the other wave numbers, the corresponding windows are smaller, i.e.\ the window
of validity for the leading-order asymptotic laws depends on the chosen correlator.
If the correction term $h_q^{(1)}G^{(1)}(t)$ in Eq.~(\ref{eq:dyn_asymptotics})
could be neglected, i.e.\ if the factorization theorem $\phi_q(t)-f_q^c=h_qG(t)$
would hold, the rescaled correlators ${\hat\phi}_q(t)=\left[\phi_q(t)-f_q^c\right]/h_q$
should collapse on the common function $G(t)$. In particular, all correlators
should cross their plateau value $f_q^c$ at the same time $t_-$, given by $G(t_-)=0$.
The latter property is fulfilled within a small error margin for $t_-=8.9\times10^5$.
Figure~\ref{fig:fig12}b demonstrates the validity of the factorization property for
a two-decade window. The size of this window is considerably smaller than the one
found for the HSS for a state with a similar overall relaxation time \cite{Franosch97}.
Thus, the next-to-leading-order correction in Eq.~(\ref{eq:dyn_asymptotics}) is much
larger near the $A_4$ than the one known from the bifurcation dynamics of the simple
HSS.

\section{Conclusions}
\label{sec:conc}

In this paper, ideal liquid-glass transitions and the evolution of glassy
dynamics were analyzed within the basic version of the mode-coupling theory (MCT)
for a simple colloid model, where the particles interact via a square-well potential.
The discussion was restricted to the high-density regime. Hence the excluded-volume
effects play a crucial role for the structure, and the cage effect is an essential
feature of the dynamics. The presence of short-ranged attractions leads to a variety
of new features compared to the ones known from the hard-sphere system (HSS).
We find a subtle phase diagram for the glass-transition lines in the plane
spanned by the two control parameters, packing fraction $\varphi$ and reduced
temperature $\theta$ (Fig.~\ref{fig:fig5}). The diagram is organized around an
$A_4$-glass-transition singularity. This occurs for a critical value
$\delta^*\approx0.04$ of the ratio $\delta$ of the attraction-well width and the
hard-core diameter, a packing fraction $\varphi^*$ exceeding the transition density
$\varphi_c^{\text{HSS}}$ of the HSS, and a certain critical temperature,
cf.\ Eq.~(\ref{eq:A4val}).

For $\delta>\delta^*$ and sufficiently low temperature,
there is a part of the liquid-glass transition line, where the critical temperature
$\theta_c$ increases with the critical density $\varphi_c$. As expected for
conventional liquids with, e.g., Lennard-Jones interactions, the glass transition
can occur either upon cooling or upon compression. This part of the transition line
extends up to densities where $\varphi$ exceeds $\varphi_c^{\text{HSS}}$, since the
bonding effects due to the attraction stabilize the liquid phase. For large
temperatures, the effects of the attraction get suppressed. Therefore, there exists a
high-temperature piece of the transition line, where $\theta_c$ decreases with
increasing $\varphi_c$. There appears a regime of high density, where the liquid
can transform to a glass either by cooling or by heating. The possibility of such a
reentry phenomenon is characteristic for systems with a hard-core repulsion. In a
conventional system, the effect cannot occur, since a soft-core repulsion implies a
decrease of the effective repulsion-core diameter with heating; and this decrease
overcompensates the effect of the decrease of the effective attraction strength.

For $\delta<\delta^*$, the two mentioned transition-line parts no longer join smoothly.
Rather the low-temperature line terminates the high-temperature one at some
crossing-point, such that they appear as two separate transition lines. At very high
temperature, the mechanism of glass formation is similar to the one of the HSS,
and in general, the temperature dependence of the high-temperature transition line
is weak. Glass transitions across this line are caused by an arrest of density
fluctuations on the length scale of the inter-particle distance. A
tagged particle is localized due to repulsion by its cage-forming neighbours.
In contrast, the low-temperature line describes glass formation due to the arrest of
density fluctuations on a length scale of the order of the attraction-shell width.
Here, tagged particles are localized due to a formation of short bonds with their
cage-forming neighbours. The density dependence of these transition points is weak,
and the transition line extends into the regime of gel formation at low densities.
On the high-density side, it extends into the glass regime, until it ends at an
$A_3$-glass-transition singularity, as indicated by the open circles in
Fig.~\ref{fig:fig5}b.

Between the mentioned line-crossing point and the endpoint of the second transition
line, there is a line of glass-glass transitions. The averaged equilibrium structure,
as characterized by the structure factor $S_q$, is the same on either side of this
line. But the two different localization mechanisms imply qualitatively different
frozen structures, reflected by differences in the Debye-Waller factor $f_q$.
The one on the high-temperature side, shown by the uppermost solid line in
Fig.~\ref{fig:fig8}, is similar to the Debye-Waller factor of the HSS at the same
density. It exhibits a pronounced peak near the position $q_0$ of the
structure-factor peak, and the zero-wave number limit $f_0^c$ is about $0.7$.
On the low-temperature side, $f_q^c$ is much larger, as is shown by the lowest of
the dashed lines in Fig.~\ref{fig:fig8}. In particular, $f_0^c$ is considerably
increased. The $f_q^c$-versus-$q$ curve for the attraction-dominated glass is
bell-shaped like a M\"o{\ss}bauer-Lamb factor. The increase of $f_0^c$ towards the
upper limit unity is connected with a large enhancement of the longitudinal
modulus. Crossing the glass-glass transition line, the longitudinal modulus
as well as the shear modulus experiences a large discontinuity, as shown in
Fig.~\ref{fig:fig9}. The large differences in the macroscopic elastic
properties are the most obvious manifestations of the two localization mechanisms
in the high-density system predicted by our theory.

\begin{figure}
\begin{center}%
\includegraphics[width=.35\textwidth,angle=270]{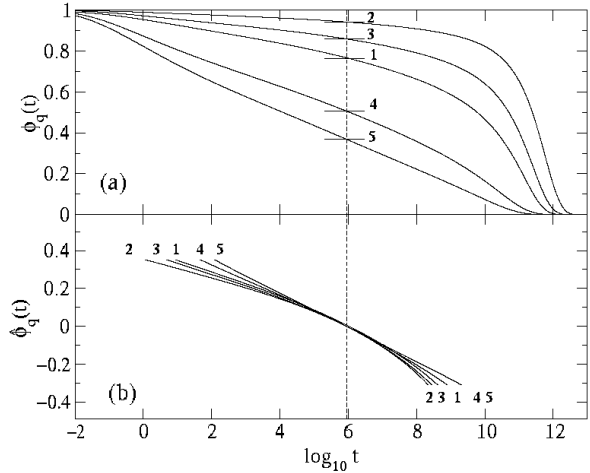}%
\end{center}
\caption{Correlators for a liquid state close to the $A_4$ singularity
calculated with the MSA for the structure factor. The curves in the upper
panel (a) show the $\phi_q(t)$ where the labels 1 to 5 indicate the wave
numbers $q=4.2,\,7.0,\,8.2,\,20.2,\,24.2$. The 
corresponding critical Debye-Waller factors $f_q^c$ are 
$0.764,\,0.943,\,0.860,\,0.507,\,0.369$, respectively. 
The curves in the lower panel (b) exhibit the rescaled correlators
${\hat\phi}_q(t) = \left[\phi_q(t) - f_q^c\right]/h_q$. Here, the critical
amplitudes $h_q = (1-f_q^c)^2 e_q$ have the values
$0.4665$, $0.1343$, $0.2881$, $0.7291$, $0.7835$.
The dashed vertical line marks the time $t_{-} = 8.9\times10^{5}$.
\label{fig:fig12}}
\end{figure}

Two general MCT predictions for the relaxation near a critical temperature or
critical density have been confirmed by many experiments and molecular-dynamics
simulations \cite{Goetze99}. First, the structural relaxation exhibits a two-step
scenario. Outside the transient, there occurs a relaxation towards the plateau $f_q^c$.
For this step, $d^2\phi_q(t)/d(\ln t)^2$ is positive. Then there is the
$\alpha$ process dealing with the relaxation from the plateau to zero. Its initial
part exhibits a negative second derivative of the $\phi_q(t)$-versus-$\ln t$
curve. Second, there holds the superposition principle for the $\alpha$ process.
On a time window that expands with increasing relaxation time, the
$\phi_q(t)$-versus-$\ln t$ curves can be collapsed on a common master curve by
shifts along the abscissa. These two simple results, which are fingerprints of the
$A_2$ bifurcation, are not valid for the relaxation at states close to an $A_4$
singularity. The curves in Fig.~\ref{fig:fig11} cannot be rescaled onto an
$\alpha$-relaxation master curve. The results in Fig.~\ref{fig:fig12} do not exhibit
changes of the second derivatives for $\phi_q(t)$ near the plateau $f_q^c$. It was
shown that the higher-order glass-transition singularities $A_3$ and $A_4$ cause
strong perturbations of the asymptotic laws usually considered, valid close to the
$A_2$ bifurcation. In the present case, they can only be observed in windows,
that might be outside the regimes accessible by experiments. In addition, the known asymptotic
laws for the relaxation near $A_3$ or $A_4$ glass-transition singularities also show up
only in windows, that are irrelevant for experimental studies. These predictions
of our theory do not seem to be a peculiarity of the square-well system. Similar
results already hold for simple one-component schematic models \cite{Sperl00}.

An exception to the findings summarized in the preceding paragraph is the
logarithmic-decay law, Eq.~(\ref{eq:logarithm}). This characteristic feature
of the dynamics near higher-order glass-transition singularities could be identified
easily in our results, as shown in Fig.~\ref{fig:fig12}. Indeed, it is shown in
Fig.~\ref{fig:fig11}, that this $\ln t$ decay is a precursor phenomenon, hindering
the evolution of the $A_2$ asymptotics. In particular, there can be
a crossover from the $\ln t$ decay to the von-Schweidler decay around the point, where
the $\phi_q(t)$-versus-$\ln t$ curve crosses the plateau $f_q^c$, as is shown by
curve 3 in Fig.~\ref{fig:fig11}. A similar scenario was recently observed for relaxation
in a micellar system \cite{Mallamace00}.

The found extreme stretching phenomena have important implications for the experimental
tests of MCT. In an experiment, it is not easily possible to measure self-averaged
correlation functions for states like the ones discussed in Fig.~\ref{fig:fig12}. Thus,
experimental results are likely to refer to history-dependent non-equilibrium states,
and ageing effects are likely to be more pronounced than they are for the normal
liquid-glass transition. Even if
proper averaging could be achieved, one cannot determine the Debye-Waller factor
$f_q=\phi_q(t\rightarrow\infty)$ within accessible time windows, if the states are
similar to the ones shown with labels 3 and 4 in Fig.~\ref{fig:fig10}. Similar
conclusions apply for the measurements of the moduli near the glass-glass-transition
line.

The presented theory is based on some assumptions which we would 
like to discuss in more detail.
First, one should expect that the equilibrium state of the system in the density regime
considered is a crystal rather than the assumed amorphous phase. In experiments
for colloids, crystallization is bypassed by chosing a polydispersity $p$ for the
particle diameters. Since nucleation rates decrease dramatically with increasing
$p$, a choice of $p$ of some percent is sufficient to establish a metastable
amorphous state for practically arbitrarily long times. A small $p$ causes only
small changes of the calculated structure factors, and thus only small changes in
the coupling coefficients entering the MCT equations. Hence a small $p$ will only imply
small changes of the presented results. Indeed, it was shown for the HSS, that a
change of $p$ did not yield detectable changes of the measured $\phi_q(t)$
\cite{Henderson96,Henderson98}. But, it is unclear how strongly, e.g., the calculated
value $\delta^*$ for the attraction-well width at the $A_4$ singularity will change,
if a realistic value for $p$ is considered.

The structure factor $S_q$ of the stable or metastable equilibrium is
used as input information for our work. Thus, the second source of reservations
is due to the errors hidden in the used $S_q$.
A well-known problem is that of the so-called thermodynamic inconsistency.
Thermodynamic quantities calculated along different routes using an approximate
$S_q$ as input often are not consistent with each other. Sophisticated closures
involving adjustable parameters could be used to overcome this
problem \cite{Hansen86}. Alas, since thermodynamics deals with the
$q\rightarrow0$ limit, for which the phase volume in the mode-coupling integrals
becomes small, one would gain no further insight carrying out our
calculations of Secs.~\ref{sec:phase} and \ref{sec:dynamics} using an improved closure
for $S_q$. For the HSS, one finds only minor changes in the numerical values for
the transition points \cite{Sperl00,Barrat89}, and the same is anticipated for the SWS.
A further difficulty arises regarding the small-$r$ limit of $c(r)$ and $g(r)$.
Due to the approximations introduced for $Q(r)$, one cannot guarantee that the
excluded-volume effect, $g(r<d)=0$, is exactly reproduced. In fact, we find that
$c(r)$ and thus $g(r)$ develop a pole $A/r$, cf.\ Eq.~(\ref{eq:cq_as}).
Since $g(r)$ is a distribution and since $A/r$ is integrable in
three dimensions, an $A/r$-term is to be viewed as small, provided $A$ is small.
In the original work on the sticky hard spheres \cite{baxter2},
$A={\mathcal O}(K^2\varphi)$. In our solution, $A={\mathcal O}(K\delta\varphi)$.
The limits $r\rightarrow0$ and $\delta\rightarrow0$ do not commute, and our analytical
solution decreases the error from a $\delta^0$ to a $\delta^1$ effect. Since our results
based on the Percus-Yevick closure and on the mean-spherical approximation are in
semi-quantitative agreement, we anticipate that better theories for $S_q$ will not
change the qualitative results of our theory.

Third, the range of applicability of the MCT is not understood.
One can use the successful tests of the theory by the experiments performed on
hard-sphere colloids, that were cited in Sec.~\ref{sec:intro}, as an \emph{a posteriori}
justification of MCT. But it is not clear, whether or not this theory can handle the
effects caused by the formation of strong short bonds. On the other hand, the phenomenon
of liquid stabilization due to bond formation and the resulting reentry effect, as well
as the drastic changes of the elastic properties at the glass-glass transition,
seem very plausible. The fact that MCT brings out these subtleties might be considered
as an argument in favour of this approach. In summary, it is the intention of this paper
to point out the possibility of new features of glassy dynamics and to suggest
a search for these features by experiments on colloids.

\acknowledgments{The work of F.S.\ and P.T.\ is supported by PRIN97-MURST 
and PRA-HOP-INFM, the work of M.F.\ by the Deutsche Forschungsgemeinschaft
grant Fu~309/3, and the work of Th.V.\ by Verbundprojekt BMBF~03-G05TUM.}

\appendix
\section{The MSA Factor Function}
\subsection{General Formulae}

Starting from Eqs.~(\ref{eq:WH1-b}) and (\ref{eq:WH2-a}),
Eq.~(\ref{eq:SWF:Qr}) for the factor function, and expressions
Eq.~(\ref{eq:SWF:abc0}) and Eq.~(\ref{eq:SWF:abc1}) for the parameters
$a, \,b,\, c$ shall be derived.
The region $0<r<1+\delta$, for which $Q(r)$ is nonzero, can be split into three parts:
\begin{equation}
\label{eq:Qr:parts}
Q(r) = \left\{
\begin{array}{ll}
q_{\text{I}}(r), 	& 0<r<\delta\\
q_{\text{II}}(r), 	& \delta<r<1\\
q_{\text{III}}(r'), 	& 0<r'=r-1<\delta
\;.
\end{array}\right.
\end{equation}
$Q(r)$ is continuous at the boundaries of the intervals, in particular
$Q(1+\delta)=0$. From Eq.~(\ref{eq:WH2-a}) together with 
$g(r)=1+h(r)$ and $G(r') = (1+r') g(1+r')$
the derivatives for the three parts of the factor function are obtained
\begin{mathletters}\label{eq:Qr:derivatives}
\begin{eqnarray}
\label{eq:Qr:derivatives:a}
q_{\text{I}}'(r)    &=& ar+b - 12\varphi\int_r^{\delta} ds\,G(s-r) q_{\text{III}}(s)\;,\\
\label{eq:Qr:derivatives:b}
q_{\text{II}}'(r)   &=& ar+b \;,\\
\label{eq:Qr:derivatives:c}
q_{\text{III}}'(r')  &=& ar+b -G(r') + 12\varphi\int_0^{r'} ds\,G(r'-s) q_{\text{I}}(s)
\;.\end{eqnarray}
\end{mathletters}
Here $g(r)=0$ for $0<r<1$ was used, and the definition for $a$ and $b$
is given in Eq.~(\ref{eq:WH:ab}).
The integrated form of  Eq.~(\ref{eq:WH1-b}) is used to introduce the
MSA closure as in Eq.~(\ref{eq:MSA}),
\begin{eqnarray}
q_{\text{III}}(r')&=&
	12\varphi\int_{r'}^{\delta} ds\,q_{\text{III}}(s) q_{\text{I}}(s-r')
\nonumber\\\label{eq:WH1-a}
&&+ K \left[1-\frac{r'}{\delta}
+\frac{\delta}{2}\left(1-\frac{r'^2}{\delta^2}\right)
\right]\;,
\end{eqnarray}
where $K=u_0\delta/k_B T$.
In the following, Eqs.~(\ref{eq:Qr:derivatives}) and (\ref{eq:WH1-a}) are
solved together with the reformulated expressions for $a$ and $b$
\begin{mathletters}\label{eq:Qr:ab_general}
\begin{eqnarray}\label{eq:Qr:ab_general:a}
a &=&	1-12\varphi\left[
\int_0^\delta ds\,q_{\text{I}}(s)
+\int_\delta^1 ds\,q_{\text{II}}(s)
+\int_0^{\delta} ds\,q_{\text{III}}(s)
\right]\;,
\\
b &=&	12\varphi\left[
\int_0^\delta ds\,s\, q_{\text{I}}(s)
+\int_\delta^1 ds\,s\, q_{\text{II}}(s)
+\int_0^{\delta} ds\,s\, q_{\text{III}}(s)\right.\nonumber\\\label{eq:Qr:ab_general:b}
&&\quad\left.
+ \int_0^{\delta} ds\,q_{\text{III}}(s)
\right]\;.
\end{eqnarray}
\end{mathletters}
Equation~(\ref{eq:Qr:derivatives:b}) gives Eq.~(\ref{eq:SWF:Qr-a}),
\begin{equation}\label{eq:qII_0sol}
q_{\text{II}}(r) = a r^2/2 + b r +c
\;, 
\end{equation}
where the continuity $q_{\text{II}}(r=1)=
q_{\text{III}}(r'=0)$ yields $c$.
For intervals I and III an expansion in $\delta$ for fixed $K$
will be performed.

\subsection{Leading Order}
In Eq.~(\ref{eq:WH1-a}), $r'/\delta$ is of order $\delta^0$ and
the integral is of higher order, $\delta^1$. 
Therefore, in leading order, 
\begin{equation}\label{eq:qIII_0sol}
q_{\text{III}}(r') = K(1-r'/\delta)\;.
\end{equation}
The boundary condition mentioned above fixes $c=K-a/2-b$ in leading order.
Substituting Eq.~(\ref{eq:qIII_0sol}) into Eq.~(\ref{eq:Qr:derivatives:c}) and 
keeping only terms in lowest order in $\delta$ results in $G(r')=K/\delta$.
The integral in Eq.~(\ref{eq:Qr:derivatives:a}) is again of higher order and the continuity
at the boundary $r=\delta$ gives the parabola also for region~I,
\begin{equation}\label{eq:qI_0sol}
q_{\text{I}}(r)=a r^2/2 + b r +c\;.
\end{equation}
Inserting the factor function into Eq.~(\ref{eq:Qr:ab_general})
and keeping only lowest order terms, linear equations for the parameters are obtained
\begin{mathletters}\label{eq:ab_eq0}
\begin{eqnarray}\label{eq:ab_eq0:a}
a &=& 1-12\varphi\left(a/6+b/2+c\right )\;,\\\label{eq:ab_eq0:b}
b &=& 12\varphi\left(a/8+b/3+c/2\right )\;,
\end{eqnarray}
\end{mathletters}
which leads to Eq.~(\ref{eq:SWF:abc0}).

\subsection{Next-To-Leading Order}
Substituting the leading order results into Eqs.~(\ref{eq:Qr:derivatives}) and (\ref{eq:WH1-a})
produces the next-to-leading order. Subtracting Eq.~(\ref{eq:Qr:derivatives:b}) from
Eq.~(\ref{eq:Qr:derivatives:a}), the result for the interval~I is given as the derivative
\begin{eqnarray}
q_{I}'(r)-q_{II}'(r)&=&-12\varphi\int_{r}^{\delta} ds\, K/\delta\, K (1-s/\delta)=
\nonumber\\\label{eq:Qr:I:1st}
&=&-6\varphi K^2\left(1-r/\delta\right)^2\;,
\end{eqnarray}
which is integrated to give the last term in Eq.~(\ref{eq:SWF:Qr-b}).
In Eq.~(\ref{eq:WH1-a}), the entire last line has to be taken into account for the 
next-to-leading order. The integral reads
\begin{eqnarray}
&&12\varphi\int_{r'}^{\delta} ds\,q_{\text{III}}(s) q_{\text{I}}(s-r')=\nonumber\\
\label{eq:Qr:III:1st}
&&\qquad = 6K\delta\cdot\varphi c_0 \left(1-r'/\delta\right)^2+ {\mathcal{O}}(\delta^2)\;.
\end{eqnarray}
Combinig Eqs.~(\ref{eq:WH1-a}) and (\ref{eq:Qr:III:1st}) yields the expression
for the next-to-leading order term for the factor function in the outer shell, 
Eq.~(\ref{eq:SWF:Qr-c}).
The continuity at $r=1$ introduces a modification of $c$ from its leading order value $c_0$,
\begin{equation}\label{eq:ab_eq1:c}
c = K-a/2-b + \delta\cdot K/2 +6\delta\cdot K \varphi c_0\;,
\end{equation}
where $a$ and $b$ are given by inserting the factor functions into Eq.~(\ref{eq:Qr:ab_general}),
\begin{mathletters}\label{eq:ab_eq1}
\begin{eqnarray}\label{eq:ab_eq1:a}
a &=& 1-12\varphi\left(a/6+b/2+c+\delta\cdot K/2\right)\;,
\\\label{eq:ab_eq1:b}
b &=& 12\varphi\left(a/8+b/3+c/2+\delta\cdot K/2\right)\;.
\end{eqnarray}
\end{mathletters}
This yields Eqs.~(\ref{eq:SWF:Qr}), (\ref{eq:SWF:abc0}), and (\ref{eq:SWF:abc1}).

\newpage

\newpage

\end{document}